\documentclass[]{jfm}

\usepackage{graphicx}
\usepackage{newtxtext}
\usepackage{newtxmath}
\usepackage{natbib}
\usepackage{hyperref}
\hypersetup{
    colorlinks = true,
    urlcolor   = blue,
    citecolor  = black,
}

\newcommand{\RomanNumeralCaps}[1]
\nolinenumbers 
\title{Boiling stratified flow: a laboratory analogy for atmospheric moist convection}

\author{Hao Fu\aff{1, 5}\corresp{\email{haofu736@gmail.com}}, Claudia Cenedese\aff{2}, Adrien Lefauve\aff{3}, and Geoffrey K. Vallis\aff{4}}

\affiliation{\aff{1}Department of the Geophysical Sciences, University of Chicago, USA
\aff{2}Physical Oceanography Department, Woods Hole Oceanographic Institution, USA
\aff{3}Department of Applied Mathematics and Theoretical Physics, University of Cambridge, UK
\aff{4}Department of Mathematics and Statistics, University of Exeter, UK
\aff{5}Department of Earth System Science, Stanford University, USA
}

\begin{document}
\maketitle

\begin{abstract}
We present a novel laboratory experiment, boiling stratified flow, as an analogy for atmospheric moist convection. A layer of diluted syrup is placed below freshwater in a beaker and heated from below. The vertical temperature profile in the experiment is analogous to the vapor mixing ratio in the atmosphere while the vertical profile of freshwater concentration in the experiment is analogous to the potential temperature profile in the atmosphere. Boiling starts when the bottom of the syrup layer reaches the boiling point, producing bubbles and vortex rings that stir the two-layer density interface and bring colder fresh water into the syrup layer. When the syrup layer at the beginning of the experiment is sufficiently thin and diluted, the vortex rings entrain more cold water than needed to remove superheating in the syrup layer, ending the boiling. When the syrup layer is deep and concentrated, the boiling is steady since the entrained colder water instantaneously removes the superheating in the bottom syrup layer. A theory is derived to predict the entrainment rate and the transition between the intermittent and steady boiling regimes, validated by experimental data. We suggest that these dynamics may share similarities with the mixing and lifecycle of cumulus convection.
\end{abstract}




\section{\label{sec:intro}Introduction}

Atmospheric convection is a particular type of fluid convection that involves phase changes. As an ascending parcel cools by adiabatic expansion, water vapor condenses and releases latent heat. This provides extra buoyancy that makes parcels penetrate the troposphere, which has a stratification stable to dry convection. As a result, in-cloud saturated parcels are unstable, but clear-sky unsaturated parcels are stable, leading to conditional instability \citep{bjerknes1938saturated}. When a cumulus cloud is deep enough, the re-evaporation of liquid water in the dry atmosphere produces a downdraft. The downdraft brings down dry air and shuts the convection. The updraft and downdraft couplet renders a convective lifecycle \citep{byers1949thunderstorm,feingold2010precipitation,dagan2018organization}.

Although observation and numerical simulation have been the main tools for studying clouds, efforts to simulate clouds in the laboratory have a long history that continues to this day. At the cloud microphysics scale, particle-laden flows with real droplets and ice interacting with turbulence have been studied in cloud chambers  \cite[e.g.,][]{shawon2021dependence}. The dynamics of the whole cloud as an entity, namely cloud dynamics, is much harder to reproduce. This is because the condensation due to lifting requires an apparatus as tall as the scale height of the saturation vapor mixing ratio (the vapor mixing ratio that makes the partial pressure of vapor reach the saturated vapor pressure, i.e., a relative humidity of 100\%), which is around 3 km \citep{wallace2006atmospheric}. To study cloud dynamics in the laboratory, analogies must therefore be found, with the awareness that no analogy is complete. There are at least three perspectives on studying the nature of moist convection.

In the first perspective, an individual cloud is considered as a buoyant plume or bubble. Possible internal sources of buoyancy include the heat released from a chemical reaction \citep{turner1963model}, gas bubbles \citep{turner1963carbonated}, heating coil \citep{narasimha2011laboratory}, and radiation from a lamp \citep{zhao2001rotating}. Although this setup cannot mimic the convective lifecycle, it suits problems where the feedback to the heat source can be neglected.

In the second perspective, moist convection is treated as a hydrodynamic instability, essentially an extension of the Rayleigh-Bénard convection problem that includes moisture \cite[e.g.,][]{benard1901tourbillons,rayleigh1916lix,chandrasekhar_hydrodynamic_2013,zhang2019moisture,fu2021linear}. Models of conditional instability qualitatively reproduced the length scale of a cloud system: with narrow ascent and a wide descent \citep{kuo1961convection,bretherton1987theory,bretherton1988field,pauluis2010idealized,vallis2019simple}. \citet{krishnamurti1998convection} designed an experiment of conditional instability using a temperature vertical gradient to generate the buoyancy stratification and a pH vertical gradient to mimic the vapor stratification. The selective radiative absorption of a pH indicator is used to mimic the latent heat release.

The third perspective is based on the realization that moist convection as a heat transfer mechanism tends to be in a quasi-equilibrium state (QE), especially in the tropics where radiative cooling balances condensation heating \citep{betts1973non,arakawa1974interaction,emanuel1994large,yano2012convective}. QE is a highly nonlinear state where conditional instability is self-regulated, and it still lacks an experimental analog. Next, we will briefly review the boundary layer quasi-equilibrium hypothesis \cite[BLQE, ][]{emanuel1989finite,raymond1995regulation}. Whether convection can occur depends on the moisture in the boundary layer. Convection induces mixing that brings up moist air and brings down dry air. As a result, convection is a ``valve" that releases the extra moisture accumulated by surface evaporation. One open question we aim to answer is: what controls the vertical mixing across the top of the boundary layer, and how does it influence the adjustment to equilibrium? Mixing can be undertaken by boundary layer convective cells \cite[e.g.,][]{lilly1968models,deardorff1970convective,arakawa1974interaction}, and the transport by cumulus updraft and downdraft. One particularly interesting question posed by \citet{thayer2015numerical} is how the cumulus updraft and downdraft produce turbulence at the top of the boundary layer and may indirectly influence mixing.

An ideal experiment of cloud dynamics should reproduce both conditional instability and QE. 
Conditional instability is a threshold behavior that requires the boundary layer to be close to saturation. Boiling reproduces such a threshold behavior, for it limits the water temperature around the boiling point by absorbing the heat of vaporization and then mixing the superheated water with the cold water above \citep{collier1994convective,oresta2009heat}. To improve the analogy to moist convection, we must also reproduce the mostly stable stratification in the atmosphere, which traps the moisture in the boundary layer and facilitates its accumulation, making convection intermittent. Our idea is thus to make a \textit{boiling stratified flow}, essentially coupling the two-layer Rayleigh-Bénard convection \citep{turner1965coupled,davaille1999two} with the boiling effect. Although some studies have investigated the boiling of two layers of immiscible fluids like water and oil \cite[e.g.,][]{mori1978configurations,mori1985classification,greene1988onset,takahashi2010experimental,filipczak2011pool,onishi2013boiling,kawanami2020liquid}, the boiling of two separate layers of \textit{miscible} fluids appears relatively novel.

\begin{figure}
\centerline{\includegraphics[width=4.7in, height=5in, keepaspectratio=true]{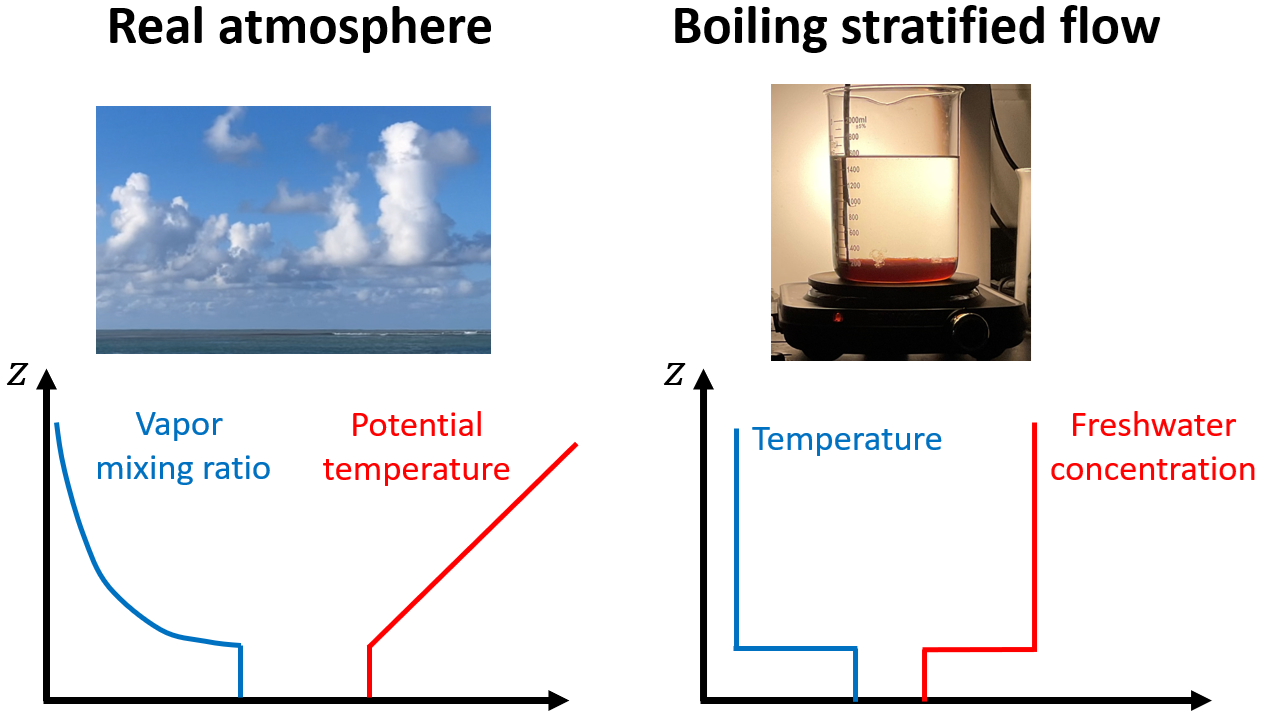}}
\caption{Schematic of the boiling stratified flow as an analogy to atmospheric convection. The left panel illustrates shallow cumulus clouds and idealised profiles of the water vapor mixing ratio and potential temperature. The right panel shows the experimental setup and the corresponding idealised profiles of temperature and freshwater concentration.} \label{fig:setup_beaker}
\end{figure}

In this paper, we propose a laboratory analogy that uses freshwater to mimic the free troposphere and a thin layer of diluted syrup to mimic the atmospheric boundary layer (Fig.\ \ref{fig:setup_beaker}). The geophysical questions we would like to address with this experiment are: 
\begin{itemize}
    \item How does moist convection favor the entrainment of dry air into the boundary layer? 
    \item How does the dry air feed back on moist convection?
\end{itemize}
The water temperature mimics the atmospheric humidity while the boiling temperature mimics the saturation vapor mixing ratio. The freshwater concentration mimics the atmospheric potential temperature, defined as the temperature of an air parcel that is adiabatically moved to the surface, representing buoyancy. Translating the questions into the context of laboratory experiments, we ask: 
 \begin{itemize}
     \item How do boiling plumes favor the entrainment of colder water into the syrup layer?
     \item How does the entrained colder water feedback on the boiling?
 \end{itemize}
Such a self-regulating behavior is analogous to the lifecycle of moist convection on Earth and giant planets \citep{feingold2010precipitation,yano2012ODEs,li2015moist}. We stress that the experiment is only an analogy, not an exact correspondence. For example, the buoyancy of a cumulus cloud is gained from the latent heat release in condensation. In the experiment, the buoyancy of a bubble is gained from the volume expansion in vaporization. This paper focuses on the fluid mechanics of boiling stratified flow and leaves a detailed comparative study with atmospheric convection for future work.

The paper is organized as follows. Section \ref{sec:setup} introduces the experimental setup. Section \ref{sec:reference} analyzes the flow evolution of the reference experiment, which inspires the theory in section \ref{sec:theory}. Section \ref{sec:sensitivity} applies the theory to understand experiments that sample the parameter space. Section \ref{sec:transition} extends the theory to study the transition between two types of boiling. Section \ref{sec:conclusion} concludes the paper. The supplemental materials, including the experimental videos, are deposited at https://doi.org/10.5281/zenodo.11222909.

\section{The experiment}\label{sec:setup}

\subsection{Experimental setup}\label{subsec:exp_setup}

The experimental setup is shown in Fig. \ref{fig:setup_beaker}. It consists of a 2000 ml beaker filled with a layer of dark corn syrup water solution underlying a thicker layer of fresh water, with a density ratio of $\rho_{s,max} /\rho_w= 1.4$, where $\rho_{s,max}\approx 1.4\times10^3$ kg m$^{-3}$ is the density of the pure syrup and $\rho_w \approx 1\times10^3$ kg m$^{-3}$ is the density of the pure water. The beaker is heated on an electric hot plate with adjustable heating power. More details on the setup are given in Appendix~\ref{app:exp_details}. When diluted, the density of the syrup solution $\rho_s$ obeys:
\begin{eqnarray}
    \rho_s = S \rho_{s,max}  + ( 1 - S ) \rho_w,
    \quad 
\end{eqnarray}
where $0\le S\le 1$ is the dimensionless syrup concentration:
\begin{eqnarray}
    S\equiv \frac{\rho_s-\rho_w}{\rho_{s,max}-\rho_w}.
\end{eqnarray}

The system is required to be statically stable at the onset of boiling. The vertical gradient of syrup concentration stabilizes the two-layer configuration against the destabilizing effect of the temperature gradient. The buoyancy $b$ (unit: m s$^{-2}$) is defined as:
\begin{eqnarray}
    b = g \left( \frac{\rho_w-\rho_s}{\rho_w} + \gamma_T T \right)
    = g \left( -\gamma_s S + \gamma_T T \right),
\end{eqnarray}
where $T$ (unit: K) is temperature, $\gamma_s \equiv (\rho_{s,max}-\rho_w)/\rho_w=0.4$ is the syrup concentration coefficient, $\gamma_T\approx 6\times10^{-4}$ K$^{-1}$ is the volumetric thermal expansion coefficient of the solution (taken as the value for 75$^\circ$C pure water, which is between the room temperature and the boiling temperature), and $g=9.8$ m s$^{-2}$ is the gravitational acceleration. At the onset of boiling, the water layer temperature is $T_w \approx 30^\circ$C (slightly above the 20$^\circ$C room temperature), while the syrup layer is around the boiling temperature $T_* \approx 100^\circ$C. We let the characteristic temperature difference be $T_*-T_w = 100^\circ\mathrm{C}-30^\circ\mathrm{C}=70^\circ\mathrm{C}$. To make the system stable, $S$ must be above a minimum value $S_{min}$:
\begin{eqnarray}  \label{eq:Smin}
    S_{min} \equiv \frac{\gamma_T \left( T_*-T_w \right)}{\gamma_s} \approx 0.11.
\end{eqnarray}

The temperature in the experiment is analogous to the water vapor mixing ratio in the atmosphere. Water vapor is not only a triggering factor of moist convection but also a component of buoyancy that makes a parcel lighter \cite[e.g.,][]{yang2018length_agg}, analogous to the temperature in the experiment that controls boiling and influences buoyancy via thermal expansion. The quantity $(1-S)$ in the experiment is the freshwater concentration. It is analogous to the potential temperature in the atmosphere, which increases with height. Because in our experiments, the initial concentration $S_0\gg S_{min}$, the buoyancy from the syrup plays the dominant role.

We use syrup because it has a higher density and viscosity than water, both of which suppress interfacial heat and mass transfer \citep{turner1986turbulent}. The high syrup viscosity is crucial in that it enables the lower layer to reach the boiling point before the two-layer stratification is eroded by turbulence (unlike what happens with a dense salt solution, as discussed in Appendix~\ref{app:exp_details}). The high viscosity does not have a direct analogy to the atmosphere, but it might be thought of as an amplifier of density stratification.

The flow is recorded with a camera from a side view, providing an integrated view of the flow field. Temperature is measured with thermocouples located at $z=1$ cm, 3 cm, 5 cm, and 7 cm above the bottom of the beaker, though this paper only discusses the $z=1$ cm and 5 cm data.

\subsection{List of experiments}

The experimenter can vary several parameters governing the system: 
\begin{enumerate}
    \item The surface heating flux $F_s$, which is analogous to the solar heating in the atmosphere.
    \item The syrup initial layer thickness $h_0$, which is analogous to the atmospheric convective boundary layer thickness.
    \item The syrup concentration $S$, which is analogous to the strength of buoyancy stratification near the top of the boundary layer. 
    \item The water layer thickness, which is analogous to the tropospheric depth.
\end{enumerate}
We decided to leave the investigation on the water layer thickness for future work by keeping the initial freshwater amount fixed to around 10 cm (corresponding to the 1400 ml scale on the beaker), much thicker than the convective penetration height. We performed four groups of experiments, varying $F_s$ (F1-F5), $h_0$ (T1-T7), $S$ (S1-S7), and $S$ and $h_0$ together (ST1-ST4), as shown in Table \ref{table_1}. Note that the labels (F3, S5), (T4, S6), (T6, ST1), and (T7, ST2) correspond to the same experiments. Experiment S3 (in bold) is the reference experiment that will be analyzed in detail. 

 We measured $S_0$ with a density meter and $h_0$ with the video (using a pixel-to-length calibration). Note that the value of $S_0$ or $h_0$  reported in Table \ref{table_1} within an experimental group (e.g., $S_0=$0.47, 0.48, 0.46, etc. for the F group) are not exactly equal due to small fluctuations introduced in preparing the two-layer fluid.

\begin{table*} 
\begin{center}
\caption{ A table of experimental parameters, which include heating voltage, surface heat flux $F_s$, initial syrup density $\rho_s$, initial syrup concentration $S_0$, and initial syrup thickness $h_0$. The post-boiling syrup thickness $h_1$ is shown in the rightmost column, with the diagnostic procedure introduced in Appendix \ref{app:diagnosis_h1}. For experiments in the steady boiling regime (section \ref{subsec:h0_sensitivity}), their $h_1$ is denoted as ``-". Note that some experiments are shared by different groups. S3 is the reference experiment (in bold). \label{table_1}}   
\begin{tabular}{ccccccc} 
\hline 
Name & Voltage (V) & $F_s$ (kW m$^{-2}$) & $\rho_s$ ($\times 10^3$ kg m$^{-3}$) & $S_0$ & $h_0$ (cm) & $h_1$ (cm)  \\ 
\hline 
F1 & 50 & 6.4 & 1.188 & 0.47 & 2.00 & 5.84 \\ 
F2 & 60 & 9.3 & 1.190 & 0.48 & 2.10 & 5.45 \\ 
F3 and S5 & 86 & 19.0 & 1.184 & 0.46 & 2.05 & 4.35 \\
F4 & 100 & 25.7 & 1.205 & 0.51 & 1.93 & 4.35 \\ 
F5 & 120 & 37.0 & 1.197 & 0.49 & 1.80 & 3.52 \\ 
\hline 
S1 & 86 & 19.0 & 1.070 & 0.18 & 1.76 & 1.98 \\
S2 & 86 & 19.0 & 1.112 & 0.28 & 1.83 & 4.74 \\
\textbf{S3} & \textbf{86} & \textbf{19.0} & \textbf{1.124} & \textbf{0.31} & \textbf{1.91} & \textbf{5.48} \\
S4 & 86 & 19.0 & 1.138 & 0.35 & 1.96 & 5.32 \\ 
S5 and F3 & 86 & 19.0 & 1.184 & 0.46 & 2.05 & 4.35 \\ 
S6 and T4 & 86 & 19.0 & 1.218 & 0.54 & 1.94 & 4.13 \\ 
S7 & 86 & 19.0 & 1.272 & 0.68 & 1.89 & 4.04 \\ 
\hline 
T1 & 86 & 19.0 & 1.205 & 0.51 &  0.80 & 4.19 \\ 
T2 & 86 & 19.0 & 1.203 & 0.51 & 1.20 & 3.90 \\ 
T3 & 86 & 19.0 & 1.199 & 0.50 & 1.50 & 4.14 \\ 
T4 and S6 & 86 & 19.0 & 1.218 & 0.54 & 1.94 & 4.13 \\ 
T5 & 86 & 19.0 & 1.211 & 0.53 & 2.64 & 4.74 \\ 
T6 and ST1 & 86 & 19.0 & 1.204 & 0.51 & 3.40 & 5.24 \\
T7 and ST2 & 86 & 19.0 & 1.204 & 0.51 &  4.13 & - \\
\hline 
ST1 and T6 & 86 & 19.0 & 1.204 & 0.51 & 3.40  & 5.24 \\
ST2 and T7 & 86 & 19.0 & 1.204 & 0.51 & 4.13  & - \\
ST3 & 86 & 19.0 & 1.136 & 0.34 & 4.08  & 6.61 \\
ST4 & 86 & 19.0 & 1.133 & 0.33 & 4.74  & - \\
\hline 
\end{tabular}
\end{center}
\end{table*}

\section{Temporal evolution of boiling stratified flow}\label{sec:reference}

\begin{figure}
\centerline{\includegraphics[width=6.3in, height=5in, keepaspectratio=true]{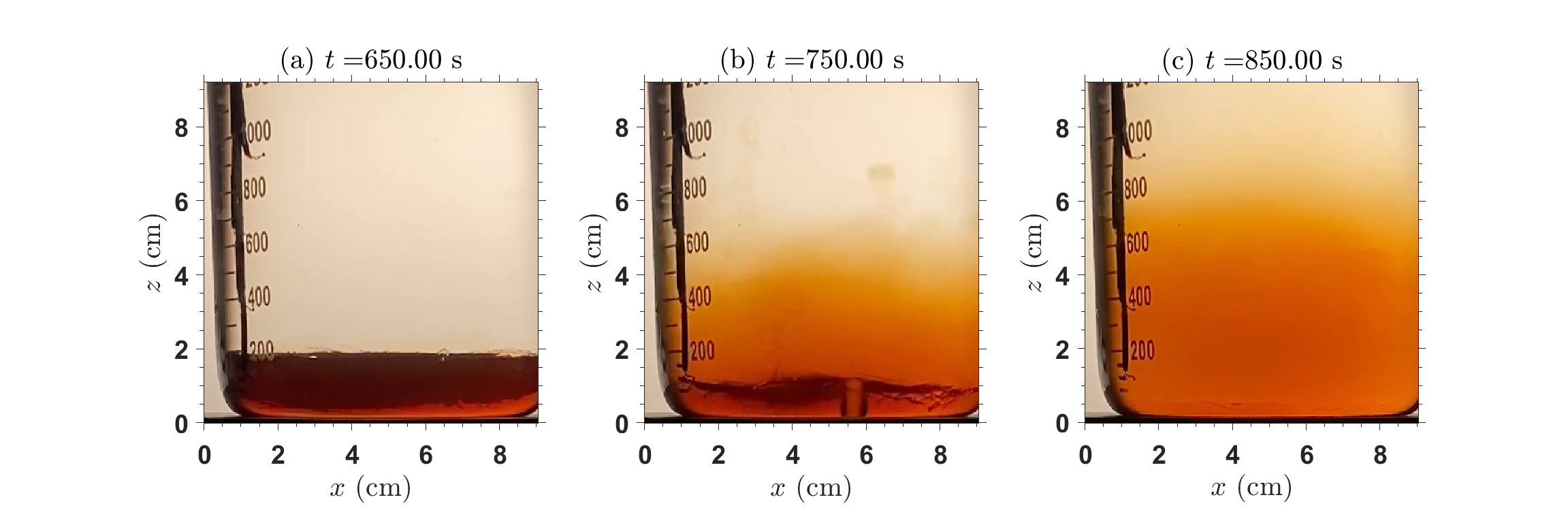}}
\caption{The flow snapshots at $t=650$ s, 750 s, and 850 s of the reference experiment (S3), showing the initial two-layer stage (a), the boiling stage (b), and the post-boiling two-layer stage (c). } \label{fig:snapshot_3}
\end{figure}

Figure \ref{fig:snapshot_3} illustrates the basic physics of boiling stratified flow in the reference experiment S3. It shows that the flow has three stages: the initial two-layer stage, the boiling stage, and the post-boiling two-layer stage. 

\begin{figure}
\centerline{\includegraphics[width=5.6in, height=5in, keepaspectratio=true]{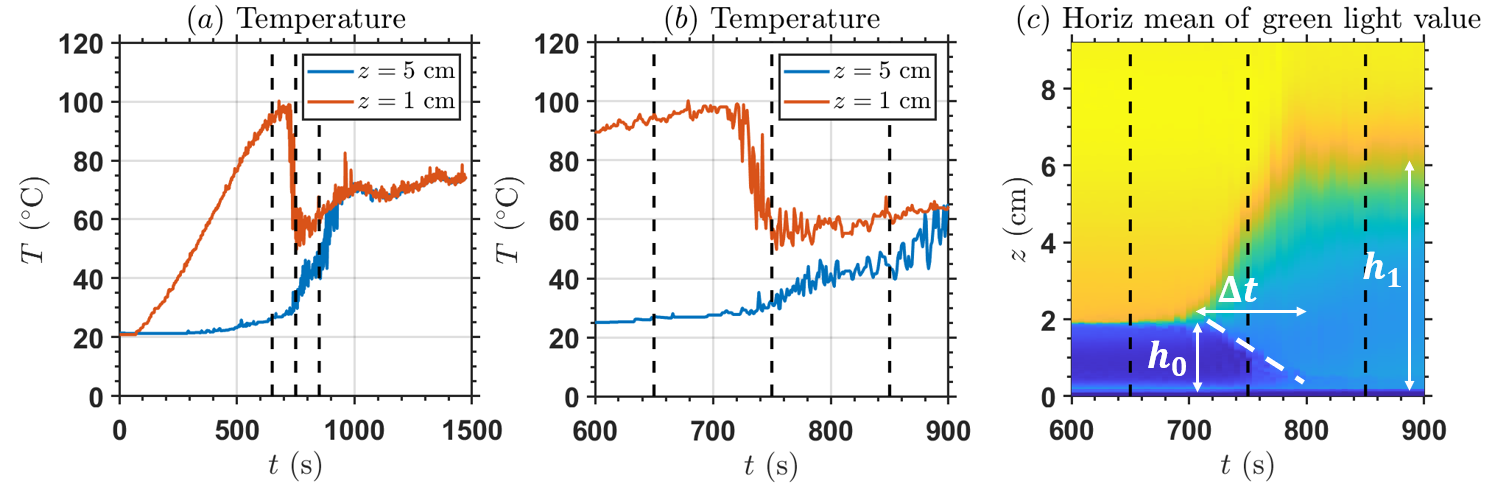}}
\caption{Quantitative measurements of the reference experiment S3. (a) The temperature time series at $z=5$ cm (blue line) and $z=1$ cm (red line). The dashed black lines denote the snapshots in Fig. \ref{fig:snapshot_3}. Note that heating starts at $t=0$. (b) The same as (a) but zooming into the boiling stage. (c) The zoom-in time evolution of the video's horizontally averaged green light pixel intensity value. The dashed white line annotates the internal interface between the bottom syrup layer and the middle mixed layer. } \label{fig:diminish}
\end{figure}

Before boiling starts, the syrup layer temperature gradually rises, and the water layer temperature remains close to the initial temperature (Fig. \ref{fig:diminish}a). This is because the density stratification suppresses heat transfer by suppressing eddy mixing \citep{turner1965coupled}. Boiling begins at around $t=700$ s by which the syrup temperature reaches 100 $^\circ$C (Fig. \ref{fig:diminish}b). Bubbles are formed near the beaker's bottom and mostly quench before leaving the syrup layer (Fig. \ref{fig:snapshot_3}b). This is because the upper part of the syrup is still below the boiling point. It can be viewed as a sub-cooled nucleate boiling phenomenon \cite[][section 4.4.4]{collier1994convective}, with the sub-cooling substantially amplified by the two-layer stratification. Though the bubble quenches, its momentum can drive a vortex ring that rises into the freshwater layer, mixes with freshwater, and sinks to the interface, producing a middle mixed layer.

Boiling only lasts about 1 minute, during which the $z=1$ cm temperature drops from 100$^{\circ}$C to around 60$^{\circ}$C and the $z=5$ cm temperature slightly rises (Fig. \ref{fig:diminish}b). The bubble-induced mixing brings colder freshwater to the syrup layer and quenches boiling. At $t=850$ s, the system still has a two-layer stratification. However, the interface rises from the initial 2 cm to 6 cm, the syrup layer is significantly diluted, and the interface is less sharp than before boiling (Fig. \ref{fig:diminish}c).  


Next, we use the horizontally averaged video pixel intensity value to track the interface height, denoted as $h$. The printed scale of the beaker is excluded from the averaging region. The video records the pixel intensity values of red, green, and blue light. We use the green light component because it captures the interface position most clearly. Figure \ref{fig:diminish}c shows boiling significantly lifts the interface from the initial value $h_0$ to a post-boiling value $h_1$. Thus, boiling can be viewed as a mixing event. We ask:
\begin{itemize}
    \item What controls the boiling duration time $\Delta t$?
    \item What controls the interface's rising rate $dh/dt$?
\end{itemize}
Knowing $\Delta t$ and $dh/dt$, we will be able to predict the net effect of mixing:
\begin{eqnarray}  \label{eq:h1_h0}
    h_1-h_0 \approx \Delta t \frac{dh}{dt}.
\end{eqnarray}

A closer look at the green light pixel intensity value (Fig. \ref{fig:diminish}c) shows an internal interface between the bottom and middle syrup layers. The internal interface splits from the outer interface at the onset of boiling and descends to the bottom at the end of boiling (dashed white line in Fig. \ref{fig:diminish}c). The vortex rings carry up syrup from the bottom syrup layer, mix with the freshwater, and deposit the mixture in the middle mixed layer. Thus, the bottom syrup layer gets thinner and finally disappears, letting the relatively cold middle mixed layer touch the bottom and quench the boiling. This suggests that $\Delta t$ is the time needed for vortex rings to remove the bottom syrup layer:
\begin{eqnarray}  \label{eq:dt}
    \Delta t \approx \frac{h_0}{\overline{w_+}},
\end{eqnarray}
where $\overline{w_+}$ (unit: m s$^{-1}$) is the horizontally averaged syrup volume flux across the internal interface, analogous to the mass flux of cumulus convection at the cloud bottom \cite[e.g.,][]{arakawa1974interaction}.

\section{Theory: from vortex rings entrainment to the final layer thickness}\label{sec:theory}

This section builds a theoretical framework to understand how $h_1-h_0$ depends on the control parameters $F_s$, $h_0$, and $S_0$. Modeling $dh/dt$ is equivalent to modeling the ensemble effect of vortex rings. 
The mixing by individual vortex rings has been investigated by \citet{olsthoorn2015vortex}. The boiling stratified flow provides a unique setup to study the nonlinear interaction between successive vortex rings.

\subsection{Escaped vs. trapped vortex rings}

\begin{figure}
\centerline{\includegraphics[width=6.9in, height=5in, keepaspectratio=true]{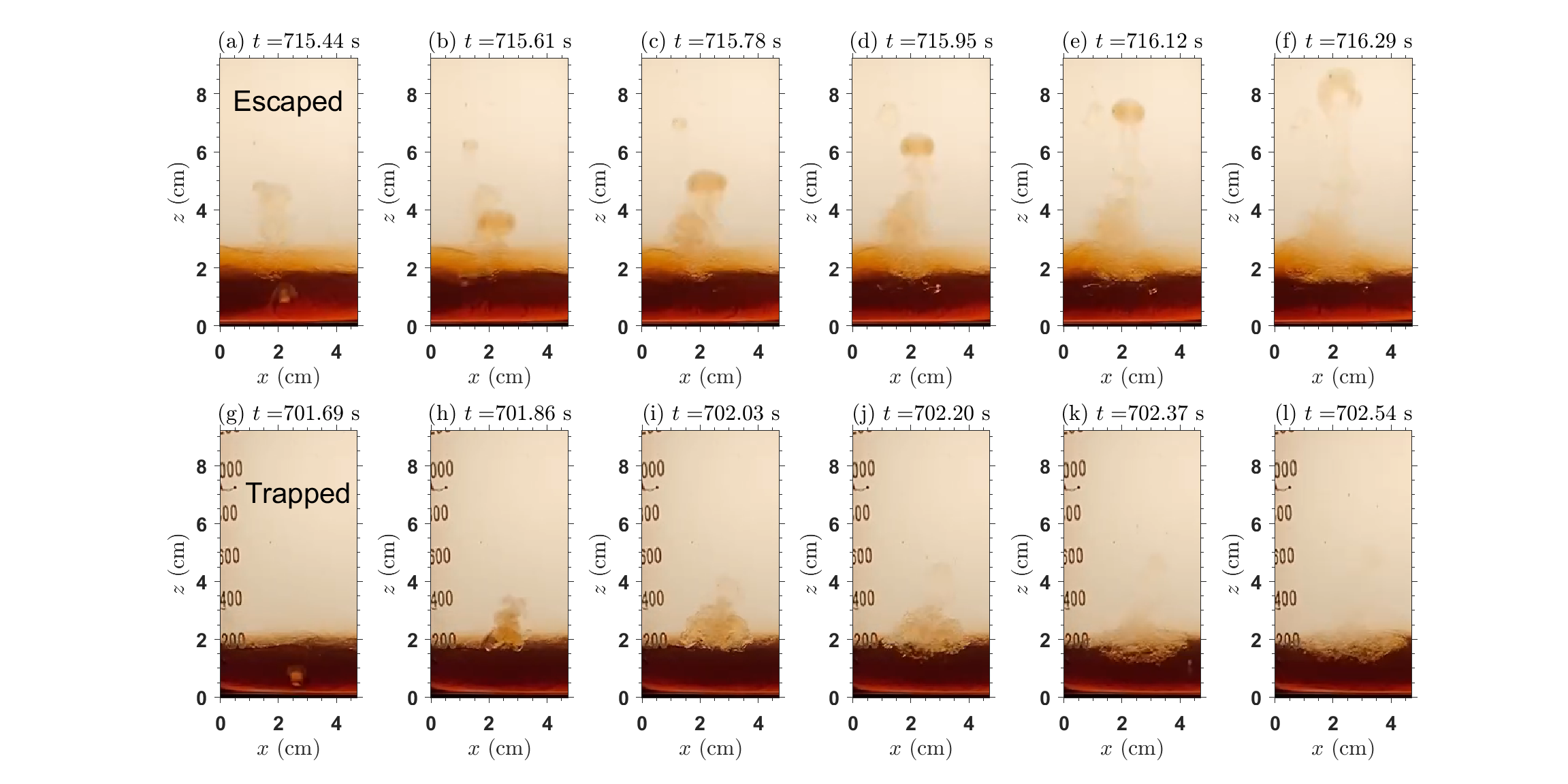}}
\caption{Examples of the two life paths of a vortex ring in the reference experiment S3. The first row shows an escaped vortex ring, and the second row shows a trapped vortex ring, both with a time interval of 0.17 s between snapshots.} \label{fig:trap}
\end{figure}

\begin{figure}
\centerline{\includegraphics[width=4.7in, height=8in, keepaspectratio=true]{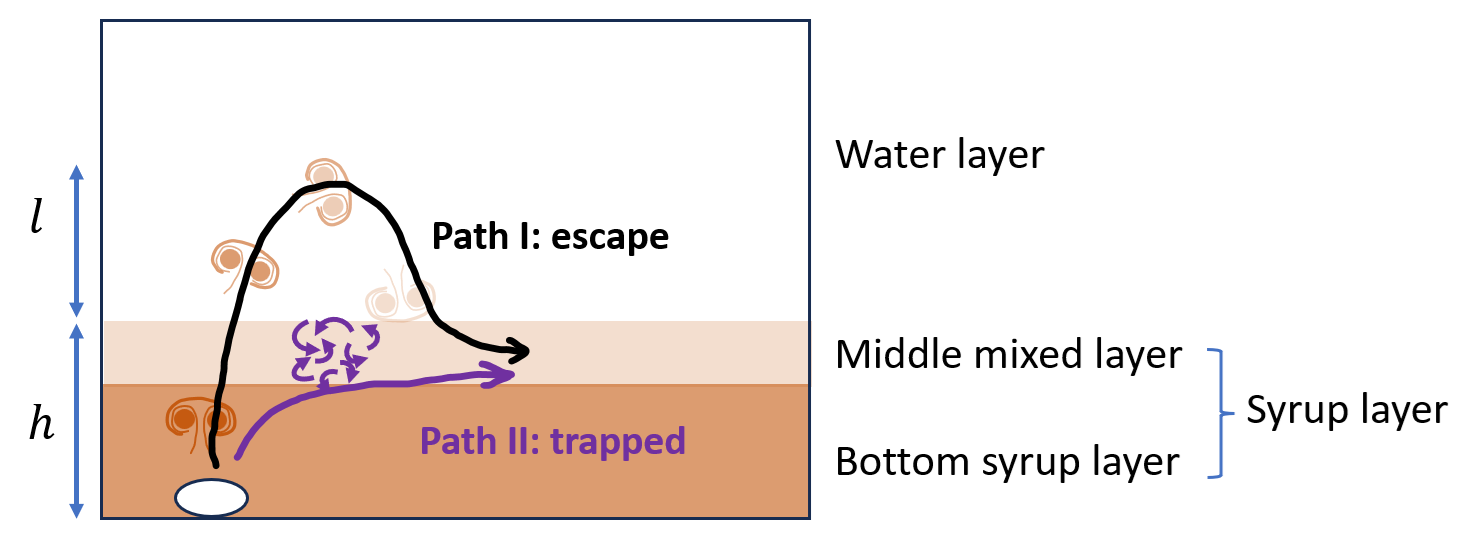}}
\caption{A schematic diagram of two life paths of a vortex ring: escaping and trapping. } \label{fig:cartoon_path}
\end{figure}

Figure \ref{fig:trap} shows the two typical life paths of vortex rings: escaping and trapping. For the escaping path, the bubble quenches in the syrup layer, leaving a vortex ring that rises into the water layer and sinks back to the interface. Note that some vortex rings completely dissipate in the freshwater layer. Their wake, consisting of syrup, can sink back to the interface. For the trapping path, the initial bubble has a similar size to the escaping case, but the vortex ring crashes near the interface, producing a wide turbulent patch. The two paths are conceptualized in Fig. \ref{fig:cartoon_path}. The escaping path has a relatively long mixing length, characterized by the vortex ring's penetration depth $l$. By contrast, the trapping path has a shorter mixing length, with a limited ability to bring down the colder water. Thus, we infer that the escaped vortex rings are mainly responsible for the thickening of the middle mixed layer.

We define the escape ratio $E$ to quantify the fraction of the vortex rings that could escape the middle mixed layer and rise into the water layer. We hypothesize that the stratified turbulence on the path of a vortex ring causes trapping. The turbulence can be induced by the wake of ascending vortex rings \citep{innocenti2021direct} or the baroclinic vorticity generated at the interface \citep{olsthoorn2018vortex}. The turbulence could tilt the orientation of the vortex ring, causing an oblique incidence onto the interface. The experiments of \citet{pinaud2021three} showed that an oblique incidence could significantly tilt the vortex ring and turn it horizontal due to the interaction between the vortex ring and the baroclinically generated vorticity at the interface. Because a smaller vortex ring is more easily tilted by turbulence, and a thicker syrup layer ($h_0$) increases the chance of tilting, we heuristically parameterize $E$ as:
\begin{eqnarray}  \label{eq:E}
    E = \exp\left( -\frac{C_E h_0}{R} \right),
\end{eqnarray}
where $C_E$ is a nondimensional escaping parameter depending on the turbulent kinetic energy in the syrup layer. In section \ref{subsec:theory_h1}, $C_E$ will be shown to be equivalent to a drag coefficient. 

We further hypothesize that $C_E$ is higher for a higher $F_s$ because stronger surface heating reduces the time interval between vortex rings. The turbulent wake of the current vortex ring could trap the next one. The trapping leads to vortex ring breaking, causing stronger turbulence and, therefore, a pileup of vortex rings. This hypothesis will be tested in section \ref{sec:sensitivity} where experiments with different $F_s$ are introduced.

\subsection{The vortex ring penetration depth $l$}\label{subsec:L}

\begin{figure}
\centerline{\includegraphics[width=4.5in, height=7in, keepaspectratio=true]{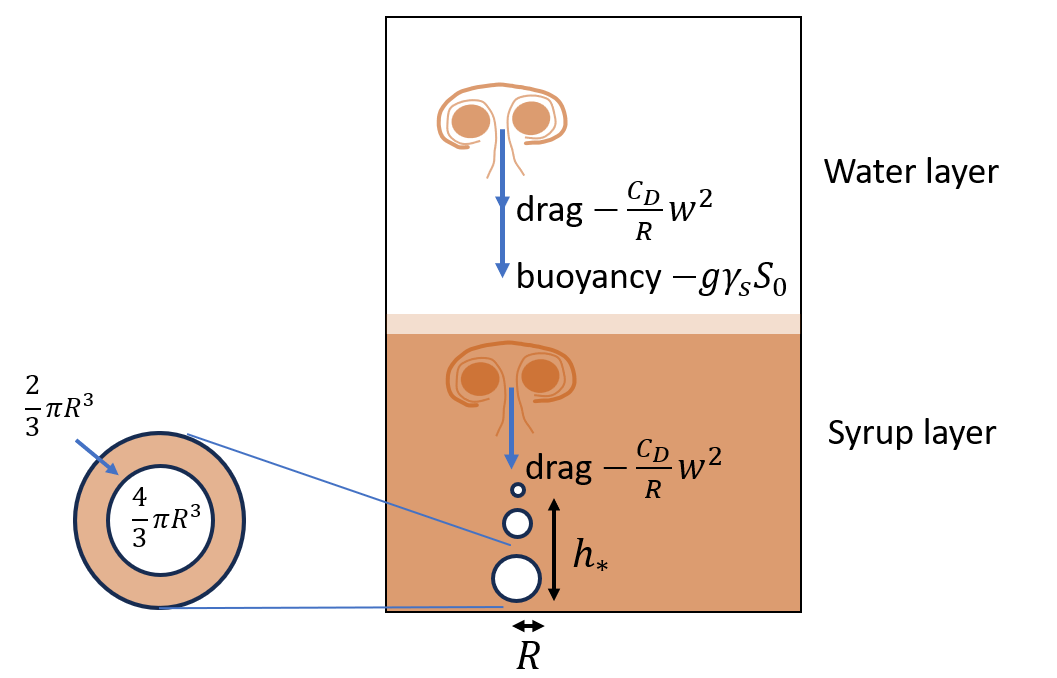}}
\caption{A schematic diagram of the vortex ring initiation and development processes. The syrup layer includes both the bottom and middle layers.} \label{fig:cartoon_force}
\end{figure}

A hotter surface temperature generally increases the initial bubble radius $R$ in boiling \citep{barathula2022review}. A bubble is highly buoyant but quenches (condenses) quickly once it leaves the hot bottom. A hotter fluid interior makes the bubble condense more slowly and yields a longer acceleration path $h_*$ for buoyancy. Combining these arguments, we see that $h_*$ should increase with the bubble radius. For simplicity, we assume:
\begin{eqnarray}  \label{eq:R_hstar}
    h_* \approx \beta R,
\end{eqnarray}
where $\beta$ is a nondimensional bubble acceleration coefficient. The bubble exerts pressure on the environment and accelerates the surrounding liquid. The added mass theory \cite[e.g.,][]{falkovich2011fluid} indicates that for a spherical bubble, the surrounding liquid moving with the bubble has half of its volume (Fig. \ref{fig:cartoon_force}). Ignoring the mass of vapor, the mean density of the bubble and the surrounding liquid is approximately $(2/3)\rho_w$. The initial velocity of the vortex ring, $w_*$, is estimated with a free-fall scaling: 
\begin{eqnarray}  \label{eq:w0}
    w_* \approx \left( 2 \frac{2}{3} g h_*  \right)^{1/2},
\end{eqnarray}
where $(2/3)g$ is the acceleration. After the bubble quenches, the moving liquid turns into a vortex ring of radius $R$.

We now perform a force analysis of an escaped vortex ring and let its vertical velocity be $w$. In the syrup layer, the vortex ring has neutral buoyancy, only influenced by drag:
\begin{eqnarray}  \label{eq:dwdt_syrup}
    \frac{dw}{dt} = - \frac{C_D}{R}w^2,
    \quad w|_{t=0} = w_*,
\end{eqnarray}
where $C_D$ is the nondimensional drag coefficient. This drag parameterization, taken from \citet{maxworthy1974turbulent}, is also used in modeling the drag of thermals in clouds \citep{romps2015sticky,romps2015stereo}. For a turbulent vortex ring, $C_D$ is a constant. For simplicity, we assume $C_D$ to be constant in our model as well. When the vortex ring enters the water layer, it is influenced by both drag and negative buoyancy:
\begin{eqnarray}  \label{eq:dwdt_freshwater}
    \frac{dw}{dt} = - \frac{C_D}{R}w^2 - g \gamma_s S_0
    \approx - g \gamma_s S_0.
\end{eqnarray}
For simplicity, we only consider the buoyancy effect after the vortex ring has escaped the syrup layer because the vortex ring would not have a maximum penetration height without considering buoyancy. The buoyancy is assumed to be controlled by $S_0$ rather than temperature, as discussed in section \ref{subsec:exp_setup}.

We can solve for the kinetic energy of the vortex ring when it crosses the interface and the negative buoyancy work done in the water layer. They are linked with the interface crossing velocity $w_+$, using (\ref{eq:dwdt_syrup}) and (\ref{eq:dwdt_freshwater}):
\begin{eqnarray}  \label{eq:KE_PE}
    \frac{2}{3} g h_* \exp \left( -\frac{2 C_D h_0}{R} \right)
    = \frac{w^2_+}{2}
    = g \gamma_s S_0 l.
\end{eqnarray}
Equation (\ref{eq:KE_PE}) yields an expression for the vortex ring penetration depth $l$:
\begin{eqnarray}  \label{eq:L}
    l = \frac{2}{3}\frac{h_*}{\gamma_s S_0} \exp \left(-\frac{2 C_D h_0}{R} \right).
\end{eqnarray}
The theory predicts that a more diluted syrup (smaller $S_0$) would make the vortex ring lighter and penetrate a longer distance. A thicker syrup layer (higher $h_0$) induces more accumulated drag and reduces $l$.

The $l$ can be used to calculate the expansion rate of the vortex ring's volume $V \equiv \frac{4}{3}\pi R^3$:
\begin{eqnarray}  \label{eq:V_R}
    \frac{1}{V} \frac{dV}{dz} 
    = \frac{3}{R} \frac{dR}{dz},
\end{eqnarray}
where $R$ is temporarily treated as a variable. For a turbulent vortex ring, the expansion rate of $R$ obeys:
\begin{eqnarray}  \label{eq:dRdz}
    \frac{dR}{dz} = \alpha.
\end{eqnarray}
Here, $\alpha$ is the nondimensional entrainment coefficient \citep{morton1956turbulent}, which is $O(10^{-2})$ \citep{maxworthy1974turbulent}. Next, we calculate the volume of the vortex ring when it sinks back to the interface, using the volume of the vortex ring when it first crosses the interface $V_+$. The traveling distance is $2l$, which includes an updraft part and a downdraft part. Further, assuming $R$ only grows by a small amount in the freshwater layer, we substitute (\ref{eq:dRdz}) into (\ref{eq:V_R}) to get:
\begin{eqnarray}  \label{eq:volume_change}
    V \approx V_+ \left( 1 + 6 \alpha \frac{l}{R} \right),
\end{eqnarray}
From now on, we return to assuming $R$ as a constant quantity. Next, we study the rising rate of the interface, which is the collective effect of many vortex rings.

\subsection{The post-boiling syrup layer thickness $h_1$}\label{subsec:theory_h1}

The rising rate $dh/dt$, where $h$ is the syrup layer thickness, depends on detrainment and entrainment across the syrup-water interface. Detrainment denotes the mass leaving the syrup layer, and entrainment denotes the mass entering the syrup layer. A vortex ring is detrained from the syrup layer first and then entrained. The interface height $h$ obeys:
\begin{eqnarray}  \label{eq:dhdt}
  \frac{dh}{dt} 
  = \underbrace{ -E \overline{w_+} }_{\text{detrain}} 
  + \underbrace{ E \overline{w_+} \frac{V}{V_+} }_{\text{entrain}}
  \approx 6 \alpha \frac{l}{R} E \overline{w_+},
\end{eqnarray}
where $\overline{w_+}$ is the horizontally averaged volume flux of vortex rings from the bottom syrup layer to the middle mixed layer, and $E\overline{w_+}$ is the flux from the middle mixed layer to the freshwater layer. The $V/V_+$ is calculated with (\ref{eq:volume_change}).

Our model is consistent with the experimental result of \citet{olsthoorn2015vortex}, who studied successive vortex rings impinging onto a stratification interface. Their experiments showed that the ratio of net entrainment to detrainment, essentially the $6 \alpha l/R$ factor in our formulation, is proportional to $\mathrm{Ri^{-1}}$. Here, $\mathrm{Ri}$ is the bulk Richardson number that obeys:
\begin{eqnarray}
    \mathrm{Ri} 
    \equiv g \frac{\rho_s - \rho_w}{\rho_w}\frac{R}{w^2_+}
    = g \gamma_s S_0 \frac{R}{w^2_+},
\end{eqnarray}
which shows $\mathrm{Ri} \propto S_0$. Substituting the expression of $l$ (\ref{eq:L}) into (\ref{eq:dhdt}), we get:
\begin{eqnarray}
    \frac{dh}{dt}
    \propto 6 \alpha \frac{l}{R} 
    \propto S_0^{-1} 
    \propto \mathrm{Ri}^{-1},
\end{eqnarray}
which is consistent with the $\mathrm{Ri}^{-1}$ scaling of their measured entrainment rate.

Combining (\ref{eq:h1_h0}), (\ref{eq:dt}), (\ref{eq:E}), (\ref{eq:L}), and (\ref{eq:dhdt}), we obtain the expression of $h_1$:
\begin{eqnarray}  \label{eq:h1_h0_full}
    h_1 = h_0 \left[ 1 + \frac{4 \alpha \beta}{\gamma_s S_0} \exp \left( - \frac{2C_D + C_E}{R} h_0 \right) \right].  
\end{eqnarray}
Note that the mean vertical volume flux of vortex rings, $\overline{w_+}$ [see Eq. (\ref{eq:dt})], does not appear in the expression of $h_1$. This expression has two unknown nondimensional parameters:
\begin{enumerate}
    \item $\alpha \beta$, the product of the entrainment parameter $\alpha$ and the bubble acceleration parameter $\beta$.
    \item The effective drag coefficients $2C_D + C_E$, which
 represents the bulk effect of the physical drag and the trapping by turbulence.
\end{enumerate}

To close the theory of $ h_1$, we still need to find an expression for the vortex ring radius $R$, which is determined by the thermodynamics of boiling.

\subsection{The bubble radius $R$}

\begin{figure}
\centerline{\includegraphics[width=4.8in, height=8in, keepaspectratio=true]{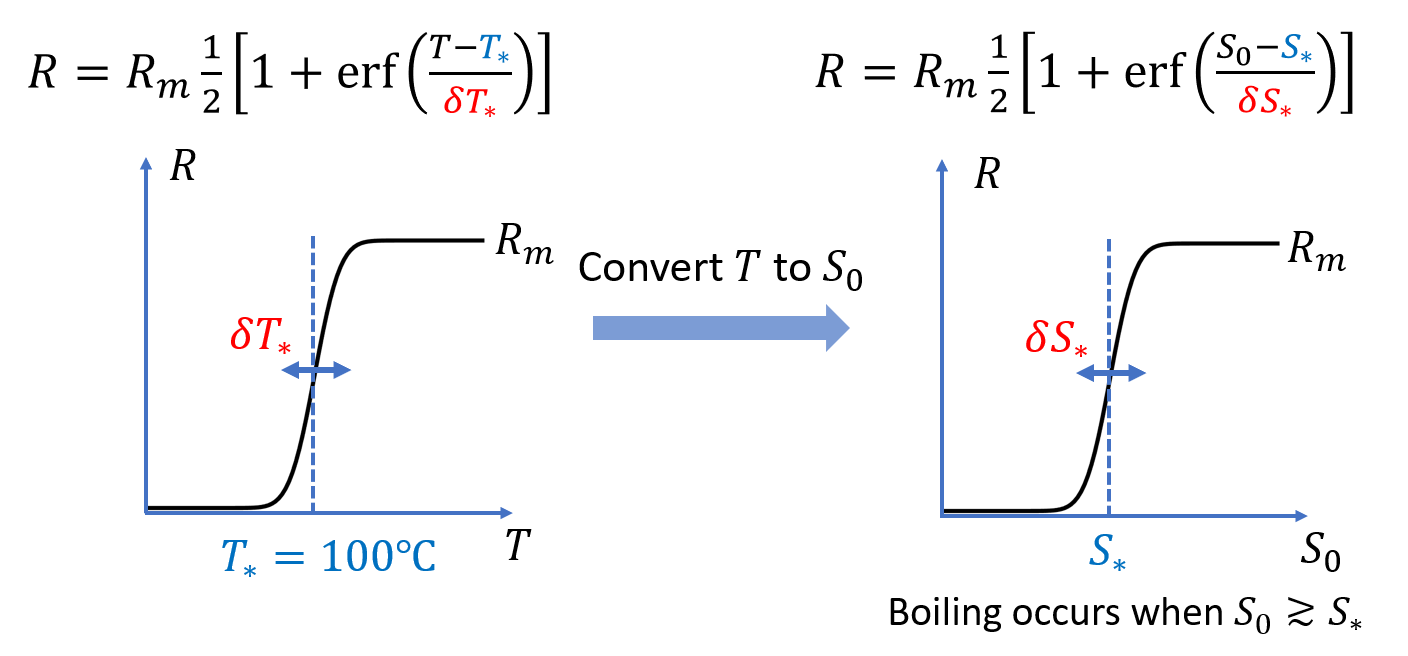}}
\caption{A schematic diagram that illustrates the parameterization of the bubble radius $R$ as a function of $T$, which is ultimately linked to the initial syrup concentration $S_0$. } \label{fig:cartoon_erf}
\end{figure}

The bubble radius $R$ in this experiment depends on how superheated the syrup layer temperature is \citep{barathula2022review}. We let the syrup layer temperature be $T$, and the boiling temperature be $T_*=100^\circ$C. When $T\ll T_*$, there is no boiling, so $R=0$. When $T\gtrsim T_*$, the water is superheated, and \citet{narayan2020experiments} showed that $R$ has an upper bound with respect to $T-T_*$, which we take as $R_m$. There is still uncertainty in the diameter-superheating relation. \citet{chang2019experimental} reported a steady increase in merged bubble diameter with the superheated temperature, though the diameter of isolated bubbles seems insensitive to superheating. See \citet{barathula2022review} for a summary. For simplicity, we acknowledge the upper bound and parameterize $R$ as an error function of $T-T_*$:
\begin{eqnarray} \label{eq:R_T}
    R \approx R_m \frac{1}{2}\left[ 1 + \mathrm{erf} \left( \frac{T-T_*}{\delta T_*} \right) \right].
\end{eqnarray}
Here, $\delta T_*$ is the temperature range of the transition zone in which $R$ is sensitive to temperature. See Fig. \ref{fig:cartoon_erf}a for an illustration.

The syrup temperature $T$ depends on the heat balance of the syrup layer, which involves the surface heat flux $F_s$, the ventilation by the interfacial heat transfer, and the vaporization and mixing caused by boiling. For simplicity, we define this equilibrium temperature without considering boiling:
\begin{eqnarray}  \label{eq:dTdt}
    \frac{dT}{dt}
    \approx 
    \underbrace{\frac{F_s}{\rho_w c_w h_0}}_{\text{surface heating}}
    - \underbrace{ \frac{T-T_w}{h_0} w_e}_{\text{ventilation}} \approx 0,
\end{eqnarray}
where $\rho_w$ is the density of pure water, $c_w$ is the specific heat of pure water [the volumetric heat capacity of $\rho_w c_w$ could approximately represent that of syrup solution, see Table A1.8 of \cite{mohos2017confectionery}], $w_e$ is the characteristic eddy vertical velocity at the syrup-water interface, and $T_w$ is the water temperature that is approximately equal to the room temperature. One might be concerned that $T$ could be unrealistically large without considering the cooling by boiling mixing. We argue this is not a serious problem because $T$ only controls the bubble radius $R$, an error function of $T$ with an upper bound. The $R$ is only sensitive to $T$ where the superheating (the surplus of liquid temperature over the boiling point) is weak and boiling is not vigorous. Thus, an overestimation of $T$ in the vigorously boiling regime should yield little error in $R$.

Equation (\ref{eq:dTdt}) shows that $T$ depends on $w_e$, with more efficient ventilation reducing $T$. Here, we model $w_e$. The syrup-layer eddy is driven by convection. It is analogous to Rayleigh-Bénard convection (RBC), with the beaker's bottom as the warm plate and the syrup-water interface as the cold plate. In our setup, the Rayleigh number (Ra) and Nusselt number (Nu) are defined as:
\begin{eqnarray}  \label{eq:Nu_Ra_definition}
    \mathrm{Ra} \equiv \frac{g\gamma_T (T-T_w) h_0^3}{\nu \kappa},
    \quad 
    \mathrm{Nu} \equiv \frac{w_e}{\kappa / h_0},
\end{eqnarray}
where $\nu$ is the kinematic viscosity and $\kappa$ is the thermal diffusivity. The Ra represents the relative strength of convective instability and the diffusive damping. The Nu represents the ratio of convective to conductive heat transfer. For the regime where the heat transfer is diffusive in the boundary layer of RBC (a thin layer attached to the beaker's bottom) and turbulent in the syrup interior, Nu obeys:
\begin{eqnarray}  \label{eq:Nu_Ra}
    \mathrm{Nu} \approx c \mathrm{Ra}^{1/3},
\end{eqnarray}
where $c=0.085$ is an empirical constant \citep{turner1965coupled}. Substituting (\ref{eq:Nu_Ra_definition}) into (\ref{eq:Nu_Ra}), we obtain an expression of $w_e$:
\begin{eqnarray}  \label{eq:wi}
    w_e 
    = c \mathrm{Ra}^{1/3} \frac{\kappa}{h_0} = c \left[ \frac{g \gamma_T (T-T_w)}{\nu \kappa} \right]^{1/3} \kappa. 
\end{eqnarray}
The $w_e$ depends on $T-T_w$, $\nu$, and $\kappa$. For syrup, $\kappa$ is insensitive to $S_0$ \cite[Table A1.9 of][]{mohos2017confectionery}. However, $\nu$ is very sensitive to $S_0$ and approximately obeys an exponential function \cite[Table A1.8 of][]{mohos2017confectionery}:
\begin{eqnarray}  \label{eq:nu}
    \nu \approx \nu_w \exp{\left(\frac{S_0}{S_{\nu}} \right)},
\end{eqnarray}
where $\nu_w$ (unit: m$^2$ s$^{-1}$) is the reference kinematic viscosity of water and $S_{\nu}$ is the critical syrup concentration for viscosity to change significantly. Substituting (\ref{eq:nu}) into (\ref{eq:wi}), we get:
\begin{eqnarray}  \label{eq:wi_viscosity}
    w_e = w_{e,ref} \exp \left( -\frac{S_0}{3 S_{\nu}}\right),
    \quad 
    w_{e,ref} \equiv c \left[ \frac{g\gamma_T (T-T_w)}{\nu_w } \right]^{1/3} \kappa^{2/3},
\end{eqnarray}
where $w_{e,ref}$ is a reference eddy velocity scale for $S_0=0$ (water). Equation (\ref{eq:wi_viscosity}) indicates that a denser syrup suppresses heat transfer.

Equation (\ref{eq:wi_viscosity}) shows the heat transfer ability of RBC, and (\ref{eq:dTdt}) shows the requirement on $w_e$ to make the syrup-layer temperature steady when neglecting boiling. Combining them and letting $T=T_*$ (the water boiling temperature 100 $^\circ$C) yields a critical $w_e$ and, therefore, a critical $S_0$ for boiling, $S_*$:
\begin{eqnarray}  \label{eq:Sstar}
    S_* = 3 S_{\nu} \ln \left[ \frac{\rho_w c_w (T_*-T_w) w_{e,ref*}}{F_s} \right].
\end{eqnarray}
where $w_{e,ref*}$ is the $w_{e,ref}$ at the boiling state:
\begin{eqnarray}
    w_{e,ref*} \equiv c \left[ \frac{g\gamma_T (T_*-T_w)}{\nu_w } \right]^{1/3} \kappa^{2/3}.
\end{eqnarray}
For $S>S_*$, convective heat transfer is too weak to keep the syrup temperature steady and below the boiling point, so boiling must occur. Substituting (\ref{eq:Sstar}) into (\ref{eq:dTdt}), we obtain the relationship between the supercritical syrup concentration ($S_0-S_*$) and the superheated temperature ($T - T_*$):
\begin{eqnarray}  \label{eq:supercritical}
    T - T_* 
    &&=  \frac{F_s}{\rho_w c_w w_{e,ref}} \exp \left(-\frac{S_0}{3S_\nu}\right) - \left( T_* - T_w \right) \nonumber \\
    &&\approx  \frac{F_s}{\rho_w c_w w_{e,ref*}} \exp \left(-\frac{S_0}{3S_\nu}\right) - \left( T_* - T_w \right) \nonumber \\    
    &&= \left[ \exp \left(\frac{S_0 - S_*}{3S_\nu}\right)  - 1 \right] \left( T_* - T_w \right)  \nonumber \\
    &&\approx \frac{S_0 - S_*}{3S_\nu} \left( T_* - T_w \right),
\end{eqnarray}
where we have used $w_{e,ref} \approx w_{e,ref*}$ in the second line and Taylor expansion in the fourth line. Equation (\ref{eq:supercritical}) indicates a denser syrup increases the superheating by increasing viscosity and suppressing heat transfer.

Substituting (\ref{eq:supercritical}) into (\ref{eq:R_T}), we express $R$ as a function of $S_0$:
\begin{eqnarray} \label{eq:R_S0}
    R = R_m \frac{1}{2}\left[ 1 + \mathrm{erf} \left( \frac{S_0-S_*}{\delta S_*} \right) \right],
    \quad 
    \delta S_* \equiv 3 S_\nu \frac{\delta T_*}{T_* - T_w} ,
\end{eqnarray}
where $\delta S_*$ is the range of the initial syrup concentration in which the bubble radius is sensitive to $S_0$ (Fig. \ref{fig:cartoon_erf}b). Equation (\ref{eq:R_S0}) indicates that a denser syrup makes bubbles larger.

Equations (\ref{eq:h1_h0_full}) and (\ref{eq:R_S0}) finally close the model for how $h_1$ depends on $h_0$ and $S_0$:
\begin{eqnarray}  \label{eq:theory_complete}
    h_1 = h_0 \left\{ 1 + \frac{4 \alpha \beta}{\gamma_s S_0} \exp \left[ - \frac{h_0}{h_{DE}}\frac{2}{ 1 + \mathrm{erf} \left( \frac{S_0-S_*}{\delta S_*} \right) } \right] \right\}.  
\end{eqnarray}
Here, $h_{DE}$ is the vortex ring dissipation length scale:
\begin{eqnarray}  \label{eq:hDE_Rm}
    h_{DE} \equiv \frac{R_m}{2C_D + C_E},
\end{eqnarray}
which represents the bulk effect of drag and trapping. The system has four unknown parameters: $\alpha \beta$, $h_{DE}$, $\delta S_*$, and $S_*$. In the next section, we apply this theory to understand the experiments with varying $F_s$, $S_0$, and $h_0$.

\section{Experimental validation of the theory}\label{sec:sensitivity}

\begin{figure}
\centerline{\includegraphics[width=7in, height=8in, keepaspectratio=true]{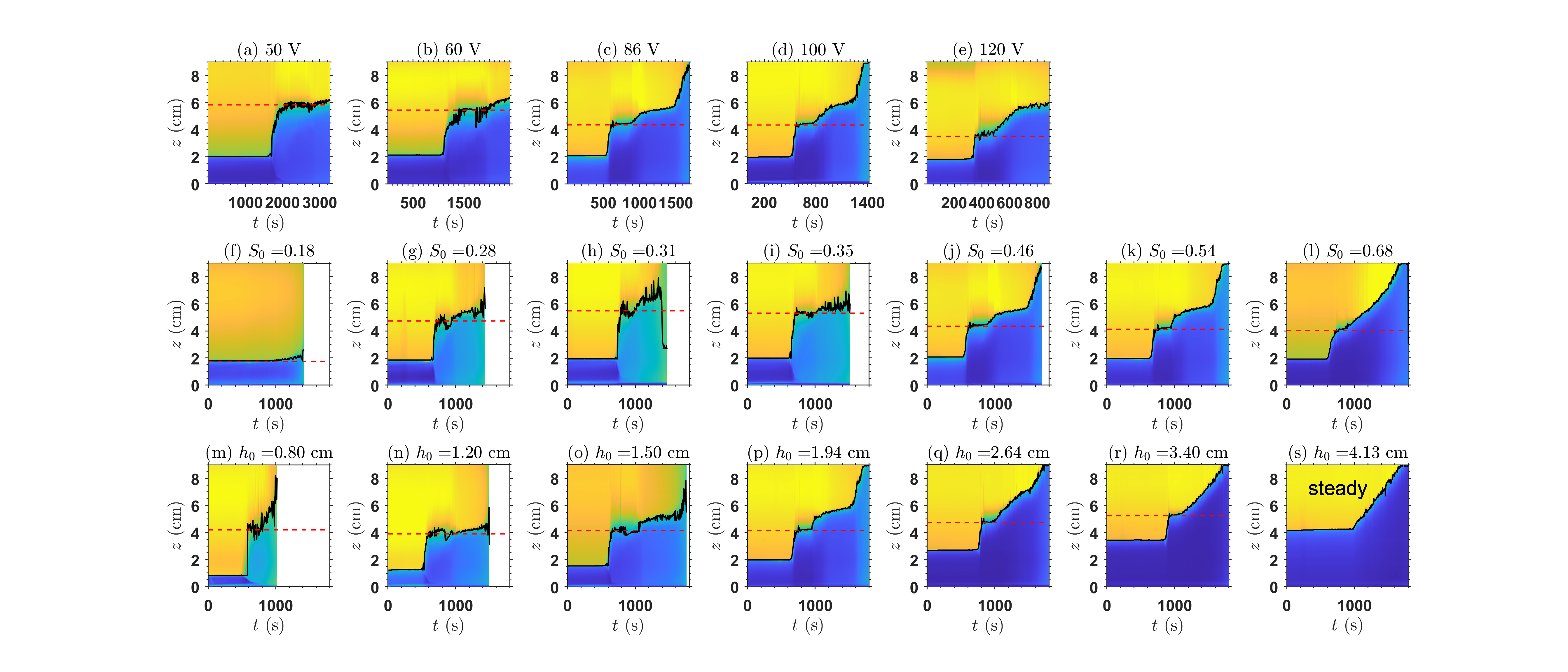}}
\caption{Time evolution of the syrup layer thickness shown with the horizontally averaged green light pixel value of the video. The first row shows experiments F1-F5, where $F_s$ is changed by varying the heating voltage. The solid black lines show the diagnosed height of the syrup-water interface. The dashed red lines show the diagnosed $h_1$. The second row is for experiments S1-S7 that change the initial syrup concentration $S_0$. The third row is for experiments T1-T7 that change the initial syrup thickness $h_0$. The T7 experiment (s) is in the steady boiling regime without a well-defined $h_1$. } \label{fig:all}
\end{figure}

To test equation (\ref{eq:theory_complete}), this section analyzes the post-boiling interface height $h_1$ diagnosed from horizontally averaged green light pixel value for experiments with varying $F_s$, $S_0$, and $h_0$.  The diagnostic method for $h_1$ (relying on imaging) is introduced in Appendix \ref{app:diagnosis_h1}, and the results are shown in Figs. \ref{fig:all} and \ref{fig:theory}.

\subsection{Experiments with varying $F_s$}

The different surface heat fluxes ($F_s$) are analogous to different solar radiative heating rates on the atmospheric lower boundary. The theory (\ref{eq:theory_complete}) shows that $F_s$ influences $h_1$ in two competing mechanisms:
\begin{enumerate}
    \item A higher $F_s$ reduces the critical syrup concentration necessary to initiate boiling, $S_*$ [Eq. (\ref{eq:Sstar})]. It should make bubbles larger and increase $h_1$.
    \item A higher $F_s$ reduces the time interval between vortex rings and increases the turbulence intensity in the syrup layer. It should increase $C_E$, trap more vortex rings, reduce entrainment, and reduce $h_1$. 
\end{enumerate}

\begin{figure}
\centerline{\includegraphics[width=6in, height=5in, keepaspectratio=true]{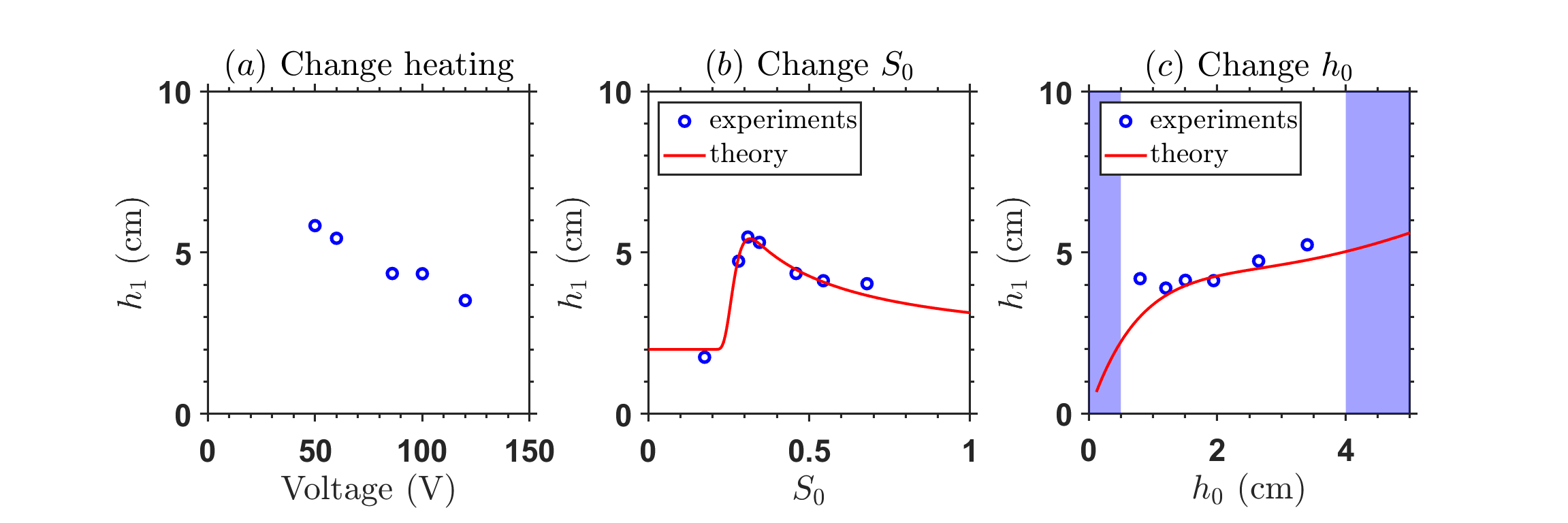}}
\caption{The post-boiling interface height $h_1$ of (a) experiments F1-F5 that vary the surface heat flux, (b) experiments S1-S7 that vary the initial syrup concentration $S_0$, and (c) experiments T1-T7 that vary the initial syrup thickness $h_0$. The blue shadings in (c) show the $h_0<0.5$ cm regime where the post-boiling state lacks a clear interface and the $h_0>4$ cm steady boiling regime. The blue dots show the experimental result, and the red lines show the theoretical prediction with the best-fit parameters. } \label{fig:theory}
\end{figure}

Experiments show that $h_1$ decreases with $F_s$ (Fig. \ref{fig:theory}a). It indicates that mechanism (ii) should play a more important role than mechanism (i). For future work, we need to answer why this is the case by quantitatively modeling how $C_E$ depends on $F_s$. It involves a careful analysis of vortex ring interaction.

\subsection{Experiments with varying $S_0$}\label{subsec:varying_S0}

The initial syrup concentration ($S_0$) is analogous to the atmospheric stratification near the boundary layer top. The theory predicts that $S_0$ influences $h_1$ with two competing mechanisms:
\begin{enumerate}
    \item A higher $S_0$ increases the viscosity. It reduces the convective ventilation of the syrup layer, enhances the superheating, and increases the vortex ring radius $R$ [Eq. (\ref{eq:R_S0})]. A higher $R$ makes the vortex ring less influenced by drag and trapping, makes the vortex ring penetrate deeper, and increases $h_1$ [Eq. (\ref{eq:theory_complete})].
    \item A higher $S_0$ makes the vortex rings more negatively buoyant, penetrate less deep, and reduce $h_1$.
\end{enumerate}

For the experimental results, $h_1$ first increases with $S_0$ and then decreases, yielding an $S_0$ around 0.3 (Fig. \ref{fig:theory}b) that maximizes $h_1$. We call it an optimal $S_0$. Thus, for the relatively dilute regime ($S_0 \lesssim 0.3$), the radius effect dominates. For the relatively dense regime ($S_0 \gtrsim 0.3$), the buoyancy effect dominates. 

\begin{figure}
\centerline{\includegraphics[width=7in, height=5in, keepaspectratio=true]{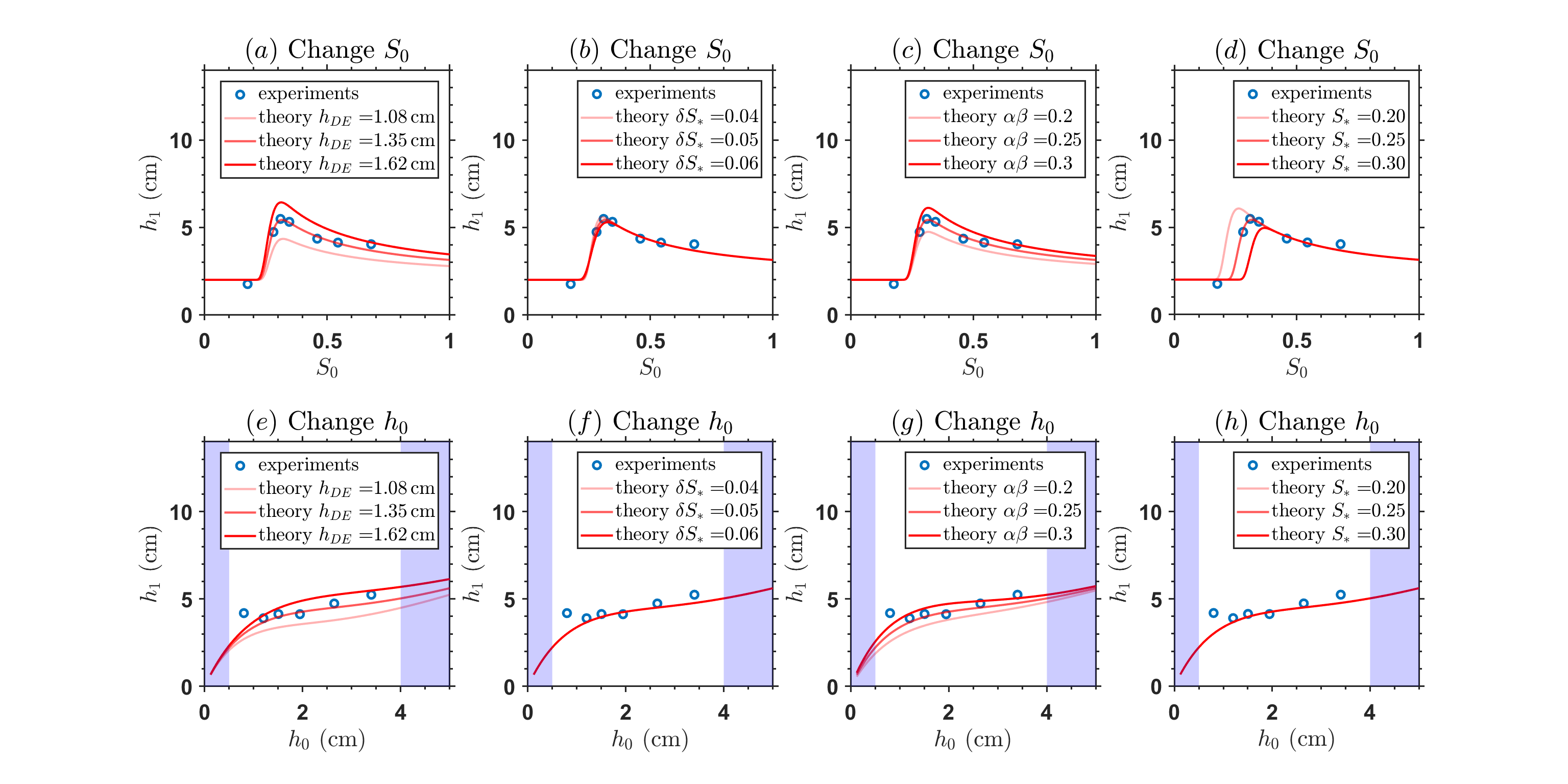}}
\caption{The first row shows the theoretical prediction of the $h_1$ vs. $S_0$ relation with perturbed parameters. The experimental results are blue circles, and the theoretical curves are solid red lines. (a) Varying $h_{DE}$. (b) Varying $\delta S_*$. (c) Varying $\alpha \beta$. (d) Varying $S_*$. The perturbation magnitude is $\pm 20\%$. The second row is the same as the first but for the $h_1$ vs. $h_0$ relation. The blue shadings show the $h_0<0.5$ cm and $h_0>4$ cm regimes where $h_1$ is not well-defined.} \label{fig:theory_sensitivity}
\end{figure}

The red lines of Fig. \ref{fig:theory}b show the quantitative prediction of $h_1$ using (\ref{eq:theory_complete}). We use $h_0=2$ cm. The values of the four unknown parameters are prescribed as $\alpha \beta=0.25$, $h_{DE}=1.35$ cm, $\delta S=0.05$, $S_* = 0.25$. This is a set of best-fit parameters, which makes the theory agree well with the experiments. The sensitivity to the values of the four parameters, perturbed with $\pm 20\%$ magnitude, is tested in the first row of Fig. \ref{fig:theory_sensitivity}, showing the trend is robust. The optimal $S_0$ mainly depends on $S_*$ and $\delta S_*$, with a higher $S_*$ and higher $\delta S_*$ shifting the optimal $S_0$ to larger values.

\subsection{Experiments with varying $h_0$} \label{subsec:h0_sensitivity}

The initial syrup thickness ($h_0$) represents the thickness of the atmospheric boundary layer. The theory predicts that $h_0$ influences $h_1$ with three competing mechanisms:
\begin{enumerate}
    \item Obviously, a higher $h_0$ makes $h_1$ higher given the same entrainment amount ($h_1-h_0$).
    \item A higher $h_0$ increases the boiling duration time $\Delta t$ because it takes longer to remove a thicker bottom syrup layer by detrainment [Eq. (\ref{eq:dt})]. This effect increases $h_1$.
    \item A higher $h_0$ increases the path for a vortex ring to be influenced by the drag and turbulence in the syrup layer, reducing its escape ratio $E$ and penetration depth $l$ [Eqs. (\ref{eq:E}) and (\ref{eq:L})]. Thus, vortex rings entrain less and should yield a lower $h_1$.
\end{enumerate}

For the experimental results, $h_1$ slightly increases with $h_0$ (Fig. \ref{fig:theory}c), indicating that the three mechanisms roughly balance each other, and a higher $h_0$ yields a less efficient dilution of the syrup layer by boiling. Using the same set of parameters as in section \ref{subsec:varying_S0} and the measured $h_0$, the trend is captured by the theory (Fig. \ref{fig:theory}c). The theoretical trend is also robust in the 20\% range sensitivity tests (Fig. \ref{fig:theory_sensitivity}). Experiments with $h_0 \lesssim 0.5$ cm yield an overly dilute post-boiling state that directly transitions to a well-mixed state. The post-boiling interface in our T1 experiment is marginally distinguishable. Experiments with $h_0 \gtrsim 4$ cm, including our T7 experiment (Fig. \ref{fig:all}s), are in a \textit{steady boiling regime} where boiling is continuous and the interfacial rising rate is steady. We name the regime with a clear end-of-boiling state the \textit{intermittent boiling regime}.

\section{Transition between the intermittent and steady boiling regimes}\label{sec:transition}

\subsection{Solving the transitional $h_0$}

We observed a steady boiling regime for a relatively high $h_0$, a regime which is not considered in the theory of section \ref{sec:theory}. Below, we extend the theory to include the steady boiling regime and study the transition between the intermittent and steady boiling regimes. 

In the steady boiling regime, the thick syrup layer sufficiently dissipates the vortex rings, reduces their penetrating depth, and limits the entrainment of the colder water. As a result, entrainment is maintained at the minimum rate that keeps the syrup temperature around $100^\circ$C:
\begin{eqnarray} \label{eq:dhdt_BLQE}
    \frac{dh}{dt} = \frac{F_s}{\rho_w c_w \left( T_* - T_w \right)}.
\end{eqnarray}
Latent heating does not appear in (\ref{eq:dhdt_BLQE}) because all bubbles condense in the syrup layer. The latent heat absorption and release balances. Using $F_s \approx 20$ kW m$^{-2}$, $\rho_w c_w \approx 4\times10^6$ J m$^{-3}$ K$^{-1}$, and $T_*-T_w=70^\circ$C, we predict a 5.7 cm rise of the interface in 800 s, which is close to the approximately 5 cm rise in 800 s shown in Fig. \ref{fig:all}s.

Next, we study what controls the transitional $h_0$. If the net entrainment rate by vortex rings [shown in Eq. (\ref{eq:dhdt})] is higher than that required to keep the syrup temperature around $100^\circ$C, boiling should be intermittent:
\begin{eqnarray}  \label{eq:intermittent_criterion}
    \mathrm{Intermittent\,\,when}:\quad
    6\,\frac{l}{R} \, E \, \overline{w_+} \, > \frac{F_s}{\rho_w c_w \left( T_* - T_w \right) },
\end{eqnarray}
where we have used (\ref{eq:dhdt}) and (\ref{eq:dhdt_BLQE}). Here, we must solve for the mean detrainment flux from the bottom syrup layer, $\overline{w_+}$, a quantity canceled out in solving $h_1$. The $\overline{w_+}$ depends on the vaporization rate at the bottom. Not all surface heating is used to vaporize water because the turbulent heat transfer between the bottom syrup layer and the middle mixed layers also cools the surface. We parameterize the ratio of vaporization cooling rate to $F_s$ as a vaporization efficiency $\chi$, an unknown parameter. We hypothesize that a lower vortex ring escape ratio $E$ enhances the turbulent mixing within the syrup layer, i.e., across the two lowest layers, and reduces $\chi$. The validation of this hypothesis is left for future work. The flux of bubble number density, $N$ (unit: m$^{-2}$ s$^{-1}$), should obey:
\begin{eqnarray}  \label{eq:N}
    N = \frac{F_s \ \chi}{\frac{4}{3}\pi R^3 L_v \rho_v },
\end{eqnarray}
where $L_v=2.5\times10^6$ J kg$^{-1}$ is the vaporization heat and $\rho_v=0.6$ kg m$^{-3}$ is the density of vapor. The added mass argument introduced in section \ref{subsec:L} indicates that the displaced volume of  syrup around a bubble is half its volume ($\frac{2}{3}\pi R^3$), hence
\begin{eqnarray}  \label{eq:w_plus}
    \overline{w_+} = \frac{2}{3}\pi R^3 N = \frac{F_s \ \chi}{2L_v \rho_v}.
\end{eqnarray}
We need to constrain $\chi$ from experiments. Figure \ref{fig:diminish}c shows that the interface between the bottom syrup layer and the middle mixed layer drops from 2 cm to 0 cm in around 60 s, indicating $\overline{w_+}\approx3.3\times10^{-4}$ m s$^{-1}$. To fit (\ref{eq:w_plus}), the $\chi$ is constrained to be $\chi = 2 L_v \rho_v \overline{w_+} / F_s \approx 0.05$, where we have used $F_s \approx 20$ kW m$^{-2}$. Thus, vaporization should play a minor role compared to turbulent mixing in cooling the syrup layer's bottom. In other words, the mechanical removal of superheating by mixing is more important than the phase change effect. Substituting the expressions for $\overline{w_+}$ (\ref{eq:w_plus}), $R$ (\ref{eq:R_S0}), $l$ (\ref{eq:L}), and $E$ (\ref{eq:E}) into (\ref{eq:intermittent_criterion}), we obtain the critical $h_0$ for transitioning to steady boiling:
\begin{eqnarray}  \label{eq:critical_h0}
    \mathrm{Steady\,\,when}:
    h_0 > h_{\perp} (S_0),\quad h_{\perp}(S_0) = h_{DE}  \ln \left[ \frac{2\alpha \beta \chi}{\gamma_s S_0} \frac{\rho_w c_w \left( T_* - T_w \right)}{L_v \rho_v} \right],
\end{eqnarray}
where $h_{\perp} (S_0)$ is the transitional $h_0$ that is a function of $S_0$. In deriving (\ref{eq:critical_h0}), we have assumed that the non-boiling regime ($S<S_*$) is sufficiently separated from the steady boiling regime by letting $R=R_m$. Substituting in estimated values ($\alpha \beta=0.25$, $\chi=0.05$, $\gamma_s=0.4$, $\rho_w c_w \approx 4\times10^6$ J m$^{-3}$ K$^{-1}$, $T_*-T_w=70^\circ$C, $L_v=2.5\times10^6$ J kg$^{-1}$, and $\rho_v=0.6$ kg m$^{-3}$), we get $h_{\perp}(S_0=0.5)=4.25$ cm. It is close to the experimental results where the critical $h_0$ lies between $h_0=3.40$ cm (T6) and $h_0=4.13$ cm (T7).

\begin{figure}
\centerline{\includegraphics[width=6.4in, height=8in, keepaspectratio=true]{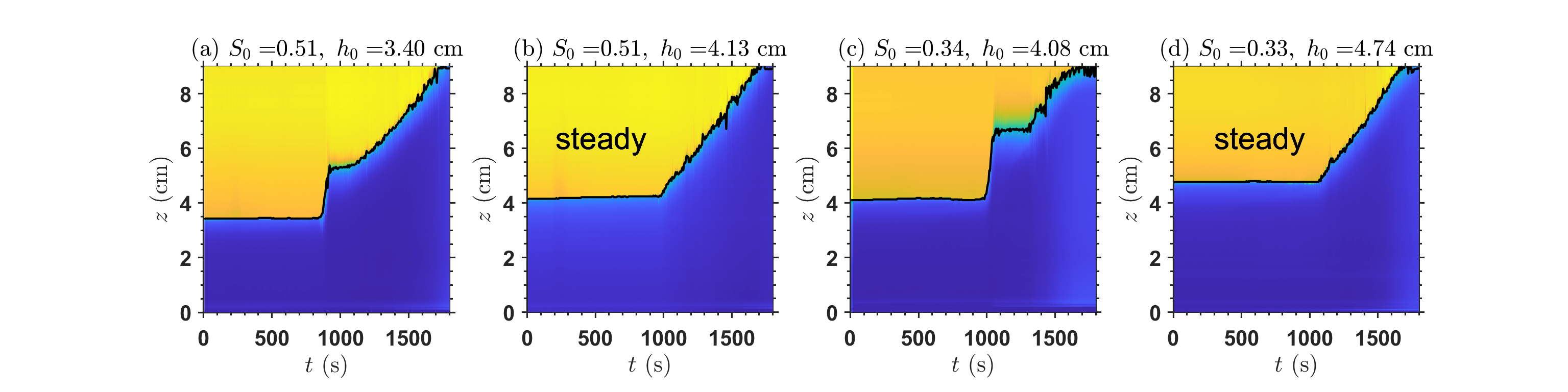}}
\caption{The same as Fig. \ref{fig:all}, but for experiments ST1-ST4 that vary both $S_0$ and $h_0$ to study the critical $h_0$. ST2 and ST4 enter the steady boiling regime at the first boiling onset. ST1 and ST3 enter it at the second onset.   } \label{fig:all_steady}
\end{figure}

Equation (\ref{eq:critical_h0}) predicts that the critical initial thickness for steady boiling $h_{\perp}$ is proportional to the dissipation length scale $h_{DE}$. The proportional factor is higher for a smaller $S_0$ because a lighter syrup penetrates deeper and entrains more. To verify the dependence of the critical $h_0$ on $S_0$, we performed experiments ST1-ST4, as shown in Fig. \ref{fig:all_steady}. For $S_0 \approx 0.5$, the critical $h_0$ lies between $h_0=3.40$ cm (ST1) and 4.13 cm (ST2). For $S_0 \approx 0.35$, the critical $h_0$ lies between $h_0=4.08$ cm (ST3) and 4.74 cm (ST4). Though 4.13 cm (ST2) is still slightly above 4.08 cm (ST3), the former is clearly in the steady boiling regime, and the latter is in the intermittent boiling regime, sufficiently showing a strong influence of $S_0$. This confirms that a more dilute syrup yields a higher critical $h_0$ for steady boiling. In summary, a smaller $h_0$ and $S_0$ increase the entrainment in boiling and make it more intermittent.

\subsection{System evolution in phase space}

The above discussions are for the first onset of boiling. We can analyze the second, third, and even $n^{th}$ onset using the same theoretical framework by taking $h_1$ as the new initial condition $h_0$. Some experiments with intermittent boiling at the first onset enter steady boiling after the syrup-layer temperature recovers to the boiling point. They include T5, T6 (ST1), and ST3, where either $S_0$ or $h_0$ is relatively large (Fig. \ref{fig:all}). For other experiments, the first onset of boiling makes the syrup layer too dilute to restore enough heat and boil again (i.e., $S<S_*$).

\begin{figure}
\centerline{\includegraphics[width=4.5in, height=8in, keepaspectratio=true]{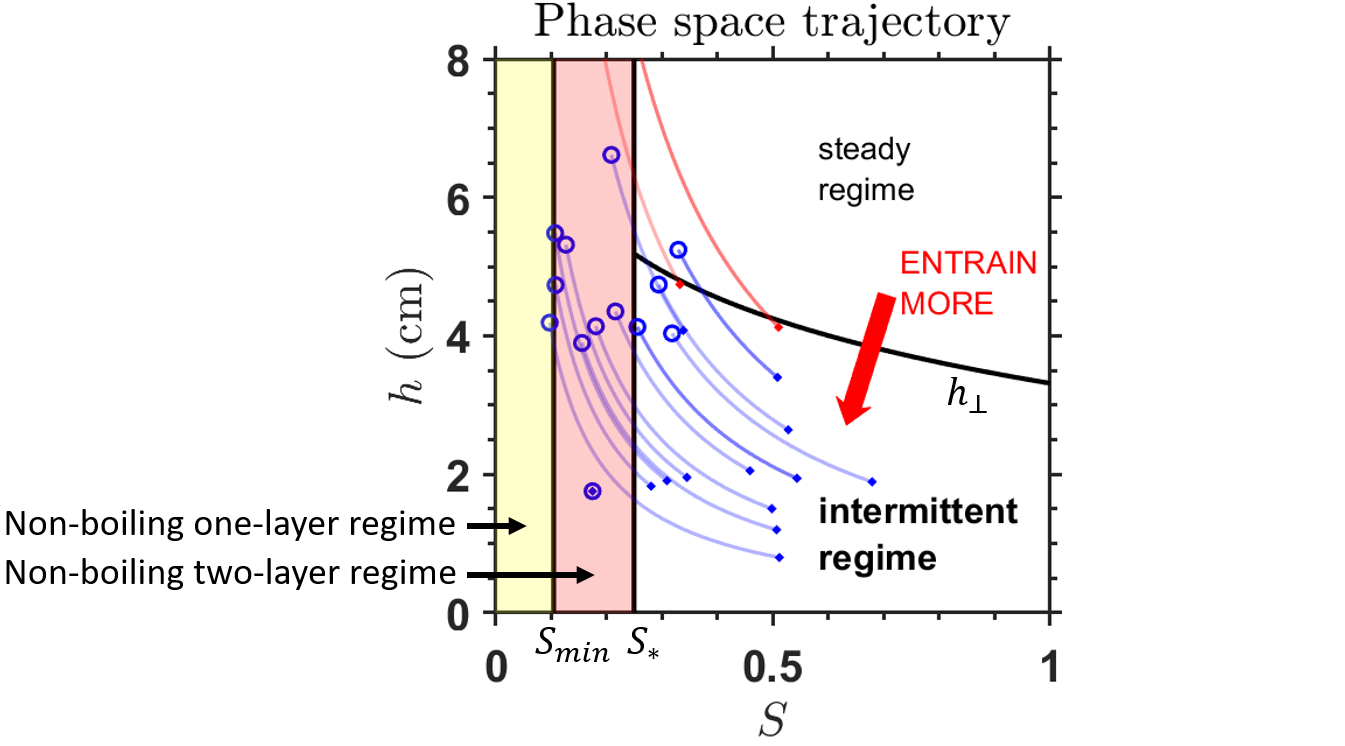}}
\caption{The system evolution in the $S$-$h$ phase space with four regimes. Phase trajectories of experiments S1-S7, T1-T7, and ST1-ST4 are plotted. Blue trajectories denote the experiments with intermittent boiling at the first onset. Red trajectories denote the experiments with steady boiling at the first onset, and no ending of the trajectory is set. The trajectories are assumed to obey $h = h_0 S_0/S$, with the dots denoting $h=h_0$ and the circles denoting $h=h_1$. Generally, a smaller $h$ or $S$ enhances the boiling entrainment. 
} \label{fig:phase_space}
\end{figure}

We summarize the system evolution in the phase space of the instantaneous syrup concentration $S$ and syrup-layer thickness $h$, as shown in Fig. \ref{fig:phase_space}. Assuming the syrup layer is diluted by entraining freshwater and no syrup is released into the freshwater layer, there should be conservation of the total syrup in the syrup layer:
\begin{eqnarray}
    S \, h = S_0 \, h_0,
\end{eqnarray}
which sets the system's trajectory in the phase space as an inverse proportional function. The parameter space is divided into four regimes:
\begin{enumerate}
    \item The non-boiling one-layer regime ($S<S_{min}$), where a two-layer configuration is convectively unstable.
    \item The non-boiling two-layer regime ($S_{min}<S<S_*$), where the convective heat transfer prevents the syrup layer from boiling.
    \item The intermittent boiling regime ($S>S_*$ and $h<h_{\perp}$).
    \item The steady boiling regime ($S>S_*$ and $h>h_{\perp}$).
\end{enumerate}
Here, the expressions of $S_{min}$ and $h_{\perp}$ are documented in (\ref{eq:Smin}) and (\ref{eq:critical_h0}).

\section{Discussion and conclusions}\label{sec:conclusion}

This paper has introduced a novel experiment, boiling stratified flow, as an analogy to vertical mixing induced by atmospheric moist convection. A thin layer of syrup in a beaker, analogous to the atmospheric boundary layer, is heated beneath a thick layer of freshwater, analogous to the free troposphere. The temperature in the experiment is analogous to the atmospheric humidity, and the threshold behavior imposed by the boiling point of the syrup solution is analogous to the saturation vapor mixing ratio.

We found that when the initial syrup concentration $S_0$ and the syrup layer thickness $h_0$ are relatively small (but $S_0$ is not too small, see Fig. \ref{fig:phase_space}), the system is in an intermittent boiling regime. The bubbles generated at the bottom of the beaker quench on their way up and drive vortex rings that penetrate the syrup-water interface, mix with water, and sink to the interface, producing a middle mixed layer that lies above the bottom syrup layer. The bottom syrup layer gradually diminishes due to the mass detrainment by vortex rings. The relatively cold middle mixed layer then contacts the bottom and ends boiling. Boiling is intermittent because more cold water is entrained into the syrup layer than is needed to remove superheating. 

We constructed a simple model of entrainment in a boiling event, which is quantified with the interface growth $h_1-h_0$. Key quantities include the escape ratio $E$, which measures the fraction of vortex rings that escape the syrup layer against the disturbance by turbulence, and the vortex ring penetration depth $l$, which determines the amount of freshwater a vortex ring can entrain. The model explains the trend of experiments with varying surface heat flux $F_s$, the initial syrup concentration $S_0$, and initial syrup layer height $h_0$. The post-boiling interface height $h_1$ decreases with increasing $F_s$ because the higher surface heating raises the bubble number density and the turbulent strength in the syrup layer and traps more vortex rings. The height $h_1$ is non-monotonic with $S_0$. For $S_0 \lesssim 0.3$, $h_1$ increases with $S_0$ due to the stronger superheating and the larger bubble radius. For $S_0 \gtrsim 0.3$, $h_1$ decreases with $S_0$ due to the more negative buoyancy of the vortex ring. The height $h_1$ is relatively insensitive to $h_0$ because a thicker syrup layer (higher $h_0$) makes a vortex ring more susceptible to drag and trapping, reducing its penetration depth $l$. This reduces $h_1-h_0$, keeping $h_0 + (h_1-h_0)$ approximately constant. We quantitatively modeled the dependence of $h_1$ on $S_0$ and $h_0$. When the four unknown parameters take the best-fit values, the agreement with experiments is very good.

When $S_0$ and $h_0$ are relatively large, the entrainment rate drops to the minimum value for removing superheating in the syrup layer. This yields a steady regime where boiling is continuous. The overshooting vortex rings continuously entrain cold water into the syrup layer. By matching the intermittent and steady regimes theory, our transition curve predicts that smaller $S_0$ raises the transitional $h_0$ to steady boiling, which qualitatively agrees with experiments. 

The experiments raised a number of questions that require further investigation, for which direct numerical simulation might be helpful. First, what determines the escape ratio $E$? How is this quantity influenced by the interaction of vortex rings and surface heating? Second, what determines the fraction of surface heating used for vaporization $\chi$? It depends on the turbulent strength in the syrup layer and may indirectly depend on $E$. These questions lead to a more fundamental question: how does the vertical eddy diffusivity change with height, and how can it be parameterized? 

Finally, let us note again that the boiling stratified flow does differ from atmospheric convection in various ways. First, the direction of phase change is the opposite of that of the atmosphere. In the experiment, buoyancy is gained by vaporization and lost by condensation. There is latent heat absorption when the bubble forms and release when the bubble quenches. Although the temperature effect on buoyancy plays a minor role compared to the bubble volume change, they make the analogy less straightforward. Second, the bubble nucleation in the experiment is different from the initiation of shallow non-precipitating cumuli in the atmosphere (Fig. \ref{fig:setup_beaker}), not to mention the more complicated precipitating deep convection. Shallow cumuli are produced by thermals in the convective boundary layer and have regular spatial patterns \cite[e.g.,][]{stull1985fair,oktem2021prediction}. There are also convective cells in the syrup layer, but bubble formation in the experiment is more likely associated with imperfections on the beaker's surface. Third, the viscosity of the syrup layer plays an important role in suppressing the interfacial heat transfer, which strengthens the sub-cooled boiling. The atmospheric boundary layer has a much higher Reynolds number than the syrup layer, and atmospheric viscosity does not play such a special role. Despite these limitations, we feel that the boiling stratified flow may be an enlightening model for studying the lifecycle and mixing process of atmospheric convection. In addition, the experiment is easy to set up and could be used as an educational demonstration to illustrate the feedbacks between convection, entrainment and stratification.

\backsection[Supplementary data]{\label{SupMat} Illustrative experimental videos, raw experimental videos, raw temperature records, and data postprocessing codes are deposited at https://doi.org/10.5281/zenodo.11222909.}

\backsection[Acknowledgements]{We thank Yunjiao Pu for providing experimental support at the kitchen stage of this project. We thank Jim McElwaine, Anders Jensen, and Bruce Sutherland for critical experimental support. We thank Quentin Kriaa, Keaton Burns, Wanying Kang, Detlef Lohse, Xi Zhang, Huazhi Ge, Zhiming Kuang, Wojciech Grabowski, Hugh Morrison, and Morgan O'Neill for stimulating discussions.}

\backsection[Funding]{We acknowledge the National Science
Foundation and the Office of Naval
Research for their support of the
2023 WHOI Geophysical Fluid Dynamics Summer Program, where the experiments were performed. AL is funded by a NERC Independent Research Fellowship  NE/W008971/1. }

\backsection[Declaration of interests]{The authors report no conflict of interest.}

\backsection[Data availability statement]{All the data is deposited in the supplemental material.}

\appendix

\section{Experimental details}\label{app:exp_details}

The beaker (model: Karter Scientific 213D20) has a volume of 2000 ml and a diameter of approximately 130 mm. In all experiments, the volume of freshwater is fixed to 1400 ml. The beaker is heated by an electric hot plate (model: SUNAV-HP102-D2, 1500 W power for 110 V voltage). The heating power is controlled by a voltage regulator (brand: VEVOR), which has a $\pm 2$ V fluctuation. The dark corn syrup (brand: Golden Barrel) has a dextrose equivalent of 42 and a density of $\rho_{s,max} \approx 1.4\times10^3$ kg m$^{-3}$. The kinematic viscosity of syrup increases approximately exponentially with its concentration [Table A1.8 of \citet{mohos2017confectionery}]. Having tried to replace the syrup layer with a much less viscous nearly saturated sodium chloride (NaCl) solution of density $1.15\times 10^3$ kg m$^{-3}$, we found that the two-layer stratification is eroded before boiling occurs.

The illumination for imaging is provided by a desk lamp diffused by a 3 mm white acrylic sheet. The light transmitted through the beaker is recorded by the camera of an iPhone 11. The temperature is recorded with K-type thermocouples (model: NUZAMAS) plugged into a temperature recorder (model: Gain Express). The temperature resolution is 0.1 $^\circ$C, and the accuracy is $\pm (1^\circ\mathrm{C}+0.3\%\mathrm{rdg})$ between 18 $^\circ$C and 28 $^\circ$C. The four sensors are bound by heat-shrink tubes and fixed to a portable retort stand.

The experimental procedure has three steps.

Step 1: Prepare the solution. First, we add approximately 1400 ml of tap water to the beaker. Then, we use an injector to add syrup to the bottom of the beaker manually. The injecting process unavoidably causes mixing and dilutes the syrup. As a rule, the injection stops when the syrup layer reaches the required $h_0$ scale on the beaker, even if the injected syrup is less than expected. Given a desired $h_0$, the volume of the syrup layer is known, but due to mixing, we use less than 100\% but typically $>75$\% of the syrup. Once the two-layer stratification is set, we use a portable resonant density meter (model: Anton Paar, DMA 35) to measure the syrup density near the bottom of the beaker.

Step 2: Start heating. We move the beaker onto the heating pad and put the temperature sensor array into it. Then, we turn on the heating pad, whose power is approximately fixed and controlled by the voltage regulator. The surface heat flux is measured via temperature rising rate in a specially designed experiment. It uses 86 V heating and 1000 ml of water, and the beaker is sealed on top by a plastic membrane to insulate heat. The surface heat flux at the bottom of the beaker is measured to be around 19 kW m$^{-2}$. The heat flux for different voltages is calculated with Ohm's law based on this reference value. 

Step 3: Cool the device. After an experiment is finished, we cool down the heating pad to be close to the room temperature before starting another experiment. An experimental cycle takes around two hours.

It is worth noticing that the experiments are sensitive to the geometry of the beaker. Ideally, we need a beaker whose bottom is not uniformly heated, which permits local superheating. The concept of superheating denotes the case where the liquid's temperature exceeds the boiling point. For our beaker, heating is strongest on a ring near the lateral boundary. This steady heating ring produces large bubbles that mix efficiently, leading to intermittent boiling. One consequence of the ring-intensified heating is that some boiling plumes are observed to crash onto the lateral wall. We have tried a 3000 ml beaker from another brand (model: ULAB, UBG1029) with a more uniform surface and an electric kettle with a perfectly uniform metal surface (model: COSORI, GK172-CO). They steadily produce tiny bubbles, a regime classified as steady boiling in section \ref{sec:transition}. Because we are particularly interested in the intermittent boiling regime with larger bubbles, which is more relevant to cumulus convection, we decided to use the beaker with a more nonuniform surface. Despite the sensitivity to the container, given the same beaker, the results are repeatable and robust.

\section{Diagnostic method of the post-boiling interface height}\label{app:diagnosis_h1}

The diagnosis of the post-boiling interface height $h_1$ has three steps. 

Step 1: Identify the syrup-water interface, essentially the top of the middle mixed layer (Fig. \ref{fig:cartoon_path}). The interface appears as a jump of the horizontally averaged pixel intensity value in the vertical direction. To facilitate the diagnosis of the jump, we vertically smooth the pixel value with a Gaussian filter whose stencil spans 20 pixels at each time snapshot. The width of a pixel, which depends on the distance of the camera to the beaker, is around 0.03 cm. The height where the vertical gradient of the smoothed pixel value is the largest is identified as the interface. This operation renders a time series of the interface height, $h(t)$, shown as the solid black line in Fig. \ref{fig:all}.

Step 2: Identify the boiling start time. We make a temporal Gaussian filter on $h(t)$ with a stencil of 20 snapshots (a time span of 101.8 s) and denote it as $\widetilde{h}(t)$. The initial value of $\widetilde{h}(t)$ is taken as $h_0$. The boiling start time is taken as the time by which $\widetilde{h}(t)$ first rises above $1.1 h_0$.

Step 3: Identify the boiling end time using $\widetilde{h}(t)$. We build a moving window spanning 20 snapshots (101.8 s) and move it from the boiling start time. The range (maximum minus minimum) of $\widetilde{h}(t)$ in the window gradually decreases as the window approaches the post-boiling stage. We let the time by which the range first drops below 0.3 cm as the boiling end time and denoted it as $h_1$. The $h_1$ is shown as the dashed red line in Fig. \ref{fig:all} and summarized in Fig. \ref{fig:theory}. The 0.3 cm threshold renders a reasonable diagnosis of $h_1$ that agrees with visual inspection. There are two exceptions. One is experiment S1 (Fig. \ref{fig:all}f), where no significant boiling occurs, and the system directly transitions to a well-mixed state after a long time (about 1400 s). The $h_1$ is taken as $h_0$. The other is experiment T7 (Fig. \ref{fig:all}s), where boiling is steady, and no boiling end time is found.

\bibliographystyle{jfm}
\bibliography{jfm}

\end{document}